\documentclass{kapproc} 
\usepackage{epsfig}
\kluwerbib

\input psfig.tex

\def\ltsima{$\; \buildrel < \over \sim \;$}
\def\simlt{\lower.5ex\hbox{\ltsima}}
\def\gtsima{$\; \buildrel > \over \sim \;$}
\def\simgt{\lower.5ex\hbox{\gtsima}}
\newcommand{\ubvri}{\protect\hbox{$U\!BV\!RI$} }
\newcommand{\bvri}{\protect\hbox{$BV\!RI$} }

\begin{document}



\articletitle{Type Ia Supernovae and Cosmology}


\author{Alexei V. Filippenko}
\affil{Department of Astronomy, University of California,\\
Berkeley, CA  94720-3411, USA} \email{alex@astro.berkeley.edu}



 \begin{abstract}
I discuss the use of Type Ia supernovae (SNe~Ia) for cosmological distance
determinations.  Low-redshift SNe~Ia ($z \simlt 0.1$) demonstrate that the
Hubble expansion is linear with $H_0 = 72 \pm 8$ km s$^{-1}$ Mpc$^{-1}$, and
that the properties of dust in other galaxies are generally similar to those of
dust in the Milky Way.  The light curves of high-redshift ($z = 0.3$--1) SNe~Ia
are stretched in a manner consistent with the expansion of space; similarly,
their spectra exhibit slower temporal evolution (by a factor of $1 + z$) than
those of nearby SNe~Ia.  The measured luminosity distances of SNe~Ia as a
function of redshift have shown that the expansion of the Universe is currently
accelerating, probably due to the presence of repulsive dark energy such as
Einstein's cosmological constant ($\Lambda$). From about 200 SNe~Ia, we find
that $H_0t_0 = 0.96 \pm 0.04$, and $\Omega_\Lambda - 1.4 \Omega_M = 0.35 \pm
0.14$. Combining our data with the results of large-scale structure surveys, we
find a best fit for $\Omega_M$ and $\Omega_\Lambda$ of 0.28 and 0.72,
respectively --- essentially identical to the recent {\it WMAP} results (and
having comparable precision). The sum of the densities, $\sim 1.0$, agrees with
extensive measurements of the cosmic microwave background radiation, including
{\it WMAP}, and coincides with the value predicted by most inflationary models
for the early Universe: the Universe is flat on large scales.  A number of
possible systematic effects (dust, supernova evolution) thus far do not seem to
eliminate the need for $\Omega_\Lambda > 0$. However, during the past few years
some very peculiar low-redshift SNe~Ia have been discovered, and we must be
mindful of possible systematic effects if such objects are more abundant at
high redshifts. Recently, analyses of SNe~Ia at $z = 1.0$--1.7 provide further
support for current acceleration, and give tentative evidence for an early
epoch of deceleration. The dynamical age of the Universe is estimated to be
$13.1 \pm 1.5$ Gyr, consistent with the ages of globular star clusters and with
the {\it WMAP} result of $13.7 \pm 0.2$ Gyr.  According to the most recent data
sets, the SN~Ia rate at $z > 1$ is several times greater than that at low
redshifts, presumably because of higher star formation rates long ago. Moreover,
the typical delay time from progenitor star formation to SN~Ia explosion
appears to be substantial, $\sim 3$~Gyr.  Current projects include the search
for additional SNe~Ia at $z > 1$ to confirm the early deceleration, and the
measurement of a few hundred SNe~Ia at $z = 0.2$--0.8 to more accurately
determine the equation-of-state parameter of the dark energy, $w = P/(\rho
c^2)$, whose value is now constrained by SNe~Ia to be in the range $-1.48
\simlt w \simlt -0.72$ at 95\% confidence.
\end{abstract}



\section*{Introduction}

    Supernovae (SNe) come in two main observational varieties (see Filippenko
1997b for a review). Those whose optical spectra exhibit hydrogen are
classified as Type II, while hydrogen-deficient SNe are designated Type
I. SNe~I are further subdivided according to the detailed appearance of the
early-time spectrum: SNe~Ia are characterized by strong absorption near
6150~\AA\ now attributed to Si~II, SNe~Ib lack this feature but instead show
prominent He~I lines, and SNe~Ic have neither the Si~II nor the He~I lines (at
least not strong ones). SNe~Ia are believed to result from the thermonuclear
disruption of carbon-oxygen white dwarfs, while SNe~II come from core collapse
in massive supergiant stars. The latter mechanism probably produces most
SNe~Ib/Ic as well, but the progenitor stars previously lost their outer layers
of hydrogen or even helium, through either winds or mass transfer onto a
companion star.

   It has long been recognized that SNe~Ia may be very useful distance
indicators for a number of reasons; see Branch \& Tammann (1992), Branch
(1998), and references therein. (1) They are exceedingly luminous, with peak
$M_B$ averaging $-19.0$ mag if $H_0 = 72$ km s$^{-1}$ Mpc$^{-1}$. (2)
``Normal'' SNe~Ia have small dispersion among their peak absolute magnitudes
($\sigma \simlt 0.3$ mag). (3) Our understanding of the progenitors and
explosion mechanism of SNe~Ia is on a reasonably firm physical basis. The
results of recent models give good fits to the observed spectra and light
curves, providing confidence that we are not far off the mark.  (4) Little
cosmic evolution is expected in the peak luminosities of SNe~Ia, and it can be
modeled.  (5) One can perform {\it local} tests of various possible
complications and evolutionary effects by comparing nearby SNe~Ia in different
environments (elliptical galaxies, bulges and disks of spirals, galaxies having
different metallicities, etc.).

   Research on SNe~Ia in the 1990s has demonstrated their enormous potential as
distance indicators. Although there are subtle effects that must indeed be
taken into account, it appears that SNe~Ia provide among the most accurate
values of $H_0$, $q_0$ (the deceleration parameter), $\Omega_M$ (the matter
density), and $\Omega_\Lambda$ [the cosmological constant, $\Lambda
c^2/(3H_0^2)$].

   For more than a decade there have been two major teams involved in the
systematic investigation of high-redshift SNe~Ia for cosmological purposes. The
``Supernova Cosmology Project'' (SCP) is led by Saul Perlmutter of the Lawrence
Berkeley Laboratory, while the ``High-Z supernova search Team'' (HZT) is led by
Brian Schmidt of the Mt. Stromlo and Siding Springs Observatories. I have been
privileged to work with both of these teams (see Filippenko 2001 for a personal
account), but my primary allegiance is now with the HZT. A few years ago, the
HZT split into two overlapping subsets: the ``Higher-Z Supernova Search Team''
led by Adam Riess of the Space Telescope Science Institute, and the ESSENCE
team (``Equation of State: SupErNovae trace Cosmic Expansion'') led by
Christopher Stubbs of Harvard University. Other groups have recently formed to
conduct similar studies, such as the supernova team of the Canada-France-Hawaii
Telescope Legacy Survey. An outgrowth of the SCP, the very large SNAP
(SuperNova/Acceleration Probe) collaboration is planning a future space
satellite that will be largely devoted to using SNe~Ia for cosmology.

\section{Homogeneity and Heterogeneity}

  Until the mid-1990s, the traditional way in which SNe~Ia were used for
cosmological distance determinations was to assume that they are perfect
``standard candles'' and to compare their observed peak brightness with that of
SNe~Ia in galaxies whose distances had been independently determined (e.g.,
with Cepheid variables). The rationale was that SNe~Ia exhibit relatively
little scatter in their peak blue luminosity ($\sigma_B \approx 0.4$--0.5 mag;
Branch \& Miller 1993), and even less if ``peculiar'' or highly reddened
objects were eliminated from consideration by using a color cut.  Moreover, the
optical spectra of SNe~Ia are usually rather homogeneous, if care is taken to
compare objects at similar times relative to maximum brightness (Riess et
al. 1997, and references therein).  Over 80\% of all SNe~Ia discovered through
the early 1990s were ``normal'' (Branch, Fisher, \& Nugent 1993).

   From a Hubble diagram constructed with unreddened, moderately distant SNe~Ia
($z \simlt 0.1$) for which peculiar motions are small and relative distances
(given by ratios of redshifts) are accurate, Vaughan et al. (1995) find that

\begin{equation}
\langle M_B({\rm max})\rangle \ = \ (-19.74 \pm 0.06) + 5\, {\rm log}\, (H_0/50)~{\rm mag}.
\end{equation}

\noindent
In a series of papers, Sandage et al. (1996) and Saha et al. (1997) combine
similar relations with {\it Hubble Space Telescope (HST)} Cepheid distances to
the host galaxies of seven SNe~Ia to derive $H_0 = 57 \pm 4$ km s$^{-1}$
Mpc$^{-1}$.

   Over the past two decades it has become clear, however, that SNe~Ia do {\it
not} constitute a perfectly homogeneous subclass (e.g., Filippenko 1997a,b).
In retrospect this should have been obvious: the Hubble diagram for SNe~Ia
exhibits scatter larger than the photometric errors, the dispersion actually
{\it rises} when reddening corrections are applied (under the assumption that
all SNe~Ia have uniform, very blue intrinsic colors at maximum; van den Bergh
\& Pazder 1992; Sandage \& Tammann 1993), and there are some significant
outliers whose anomalous magnitudes cannot be explained by extinction alone.

    Spectroscopic and photometric peculiarities have been noted with increasing
frequency in well-observed SNe~Ia. A striking case is SN 1991T; its pre-maximum
spectrum did not exhibit Si~II or Ca~II absorption lines, yet two months past
maximum brightness the spectrum was nearly indistinguishable from that of a
classical SN~Ia (Filippenko et al. 1992b; Ruiz-Lapuente et al. 1992; Phillips
et al. 1992).  The light curves of SN 1991T were slightly broader than the
SN~Ia template curves, and the object was probably somewhat more luminous than
average at maximum. Another well-observed, peculiar SNe~Ia is SN 1991bg
(Filippenko et al. 1992a; Leibundgut et al. 1993; Turatto et al. 1996).  At
maximum brightness it was subluminous by 1.6 mag in $V$ and 2.5 mag in $B$, its
colors were intrinsically red, and its spectrum was peculiar (with a deep
absorption trough due to Ti~II).  Moreover, the decline from maximum was very
steep, the $I$-band light curve did not exhibit a secondary maximum like normal
SNe~Ia, and the velocity of the ejecta was unusually low. The photometric
heterogeneity among SNe~Ia is well demonstrated by Suntzeff (1996) with objects
having excellent \bvri light curves.

\section{Cosmological Uses: Low Redshifts}

   Although SNe~Ia can no longer be considered perfect ``standard candles,''
they are still exceptionally useful for cosmological distance
determinations. Excluding those of low luminosity (which are hard to find,
especially at large distances), most of the nearby SNe~Ia that had been
discovered through the early 1990s were {\it nearly} standard (Branch et
al. 1993; but see Li et al. 2001b for more recent evidence of a higher
intrinsic peculiarity rate).  Also, after many tenuous suggestions (e.g.,
Pskovskii 1977, 1984; Branch 1981), Phillips (1993) found convincing evidence
for a correlation between light-curve shape and the luminosity at maximum
brightness by quantifying the photometric differences among a set of nine
well-observed SNe~Ia, using a parameter [$\Delta m_{15}(B)$] that measures the
total drop (in $B$ magnitudes) from $B$-band maximum to $t = 15$ days later. In
all cases the host galaxies of his SNe~Ia have accurate relative distances from
surface brightness fluctuations or from the Tully-Fisher relation.  The
intrinsically bright SNe~Ia clearly decline more slowly than dim ones, but the
correlation is stronger in $B$ than in $V$ or $I$.

   Using SNe~Ia discovered during the Cal\'an/Tololo survey ($z \simlt 0.1$),
Hamuy et al. (1995, 1996b) refine the Phillips (1993) correlation between peak
luminosity and $\Delta m_{15}(B)$. Apparently the slope is steep only at low
luminosities; thus, objects such as SN 1991bg skew the slope of the
best-fitting single straight line. Hamuy et al. reduce the scatter in the
Hubble diagram of normal, unreddened SNe~Ia to only 0.17 mag in $B$ and 0.14
mag in $V$; see also Tripp (1997). Yet another parameterization is the
``stretch'' method of Perlmutter et al. (1997) and Goldhaber et al. (2001): the
$B$-band light curves of SNe~Ia appear nearly identical when expanded or
contracted temporally by a factor $(1+s)$, where the value of $s$ varies among
objects. In a similar but distinct effort, Riess, Press,
\& Kirshner (1995) show that the luminosity of SNe~Ia correlates with
the detailed {\it shape} of the overall light curve.

   By using light-curve shapes measured through several different filters,
Riess, Press, \& Kirshner (1996a) extend their analysis and objectively
eliminate the effects of interstellar extinction (as is also now done with the
$\Delta m_{15}$ method; Phillips et al. 1999).  A SN~Ia that has an unusually
red $B-V$ color at maximum brightness is assumed to be {\it intrinsically}
subluminous if its light curves rise and decline quickly, or of normal
luminosity but significantly {\it reddened} if its light curves rise and
decline more slowly. With a set of 20 SNe~Ia from the Cal\'an/Tololo sample
(Hamuy et al. 1996c) and the Harvard/Smithsonian Center for Astrophysics (CfA)
follow-up program (Riess et al. 1999a), Riess et al. (1996a) show that the
dispersion decreases from 0.52 mag to 0.12 mag after application of this
``multi-color light-curve shape'' (MLCS) method. The results from an expanded
set of nearly 50 SNe~Ia indicate that the dispersion decreases from 0.44 mag to
0.15 mag (Figure 1). The resulting Hubble constant is $65 \pm 2$ (statistical)
$\pm 7$ (systematic) km s$^{-1}$ Mpc$^{-1}$, with an additional systematic and
zero-point uncertainty of $\pm 5$ km s$^{-1}$ Mpc$^{-1}$. (Re-calibrations of
the Cepheid distance scale, and other recent refinements, lead to a best
estimate of $H_0 = 72 \pm 8$ km s$^{-1}$ Mpc$^{-1}$, where the error bar
includes both statistical and systematic uncertainties; Freedman et al. 2001.)
Saha et al. (2001) still argue for a longer distance scale, with $H_0 \approx
60$ km s$^{-1}$ Mpc$^{-1}$, using different choices in measuring the supernova
distances (Parodi et al. 2000) and the Cepheid distances; see the discussions
by Gibson et al. (2000) and Jha (2002).

   Apart from the controversy regarding their true peak luminosity,
low-redshift SNe~Ia provide the best evidence that the Hubble flow is linear
(Riess et al. 1996a), with deviations from linearity well-explained by random
peculiar velocities and the local flow field (Riess et al. 1997; Jha
2002). Phillips et al. (1999) and Riess et al. (1996a) further argue that the
dust affecting SNe~Ia is {\it not} of circumstellar origin, and show
quantitatively that the extinction curve in external galaxies typically does
not differ from that in the Milky Way (cf. Branch \& Tammann 1992, but see
Tripp 1998).

\medskip

\hbox{
\hskip +0.4truein
\vbox{\hsize 3.3 truein
\psfig{figure=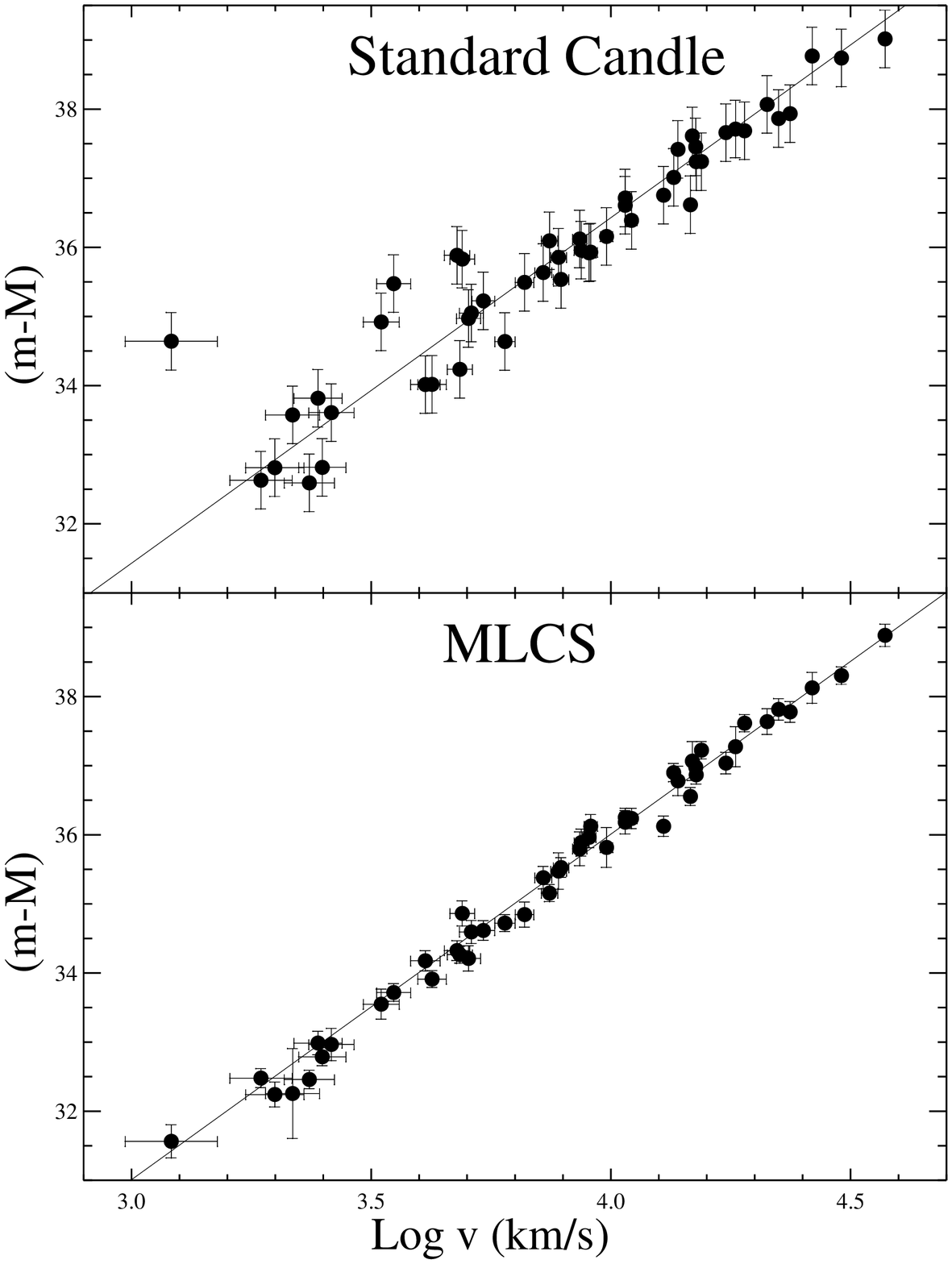,height=4.0truein,angle=0}
}
}

\bigskip
\bigskip

\noindent
{\it Figure 1:} Hubble diagrams for SNe~Ia (A. G. Riess 2001, private
communication) with velocities (km s$^{-1}$) in the {\it COBE}\ rest frame on
the Cepheid distance scale. The ordinate shows distance modulus, $m - M$ (mag).
{\it Top:} The objects are assumed to be {\it standard candles} and there is no
correction for extinction; the result is $\sigma = 0.42$ mag and $H_0 = 58 \pm
8$ km s$^{-1}$ Mpc$^{-1}$. {\it Bottom:} The same objects, after correction for
reddening and intrinsic differences in luminosity. The result is $\sigma =
0.15$ mag and $H_0 = 65 \pm 2$ (statistical) km s$^{-1}$ Mpc$^{-1}$, subject to
changes in the zero-point of the Cepheid distance scale. (Indeed, the latest
results with SNe~Ia favor $H_0 = 72$ km s$^{-1}$ Mpc$^{-1}$.)

\bigskip

   The advantage of systematically correcting the luminosities of SNe~Ia at
high redshifts rather than trying to isolate ``normal'' ones seems clear in
view of evidence that the luminosity of SNe~Ia may be a function of stellar
population. If the most luminous SNe~Ia occur in young stellar populations
(e.g., Hamuy et al. 1996a, 2000; Branch, Romanishin, \& Baron 1996; Ivanov,
Hamuy, \& Pinto 2000), then we might expect the mean peak luminosity of
high-$z$ SNe~Ia to differ from that of a local sample. Alternatively, the use
of Cepheids (Population I objects) to calibrate local SNe~Ia can lead to a
zero-point that is too luminous.  On the other hand, as long as the physics of
SNe~Ia is essentially the same in young stellar populations locally and at high
redshift, we should be able to adopt the luminosity correction methods
(photometric and spectroscopic) found from detailed studies of low-$z$ SNe~Ia.

   In the past few years, many nearby SNe have been found by industrious
amateur astronomers including R. Arbour, M. Armstrong, T. Boles, T. Puckett,
M. Schwartz, and others. The ``Nearby Supernova Factory'' run by G. Aldering's
team at the Lawrence Berkeley National Laboratory is also responsible for many
discoveries, when it is conducting a search. However, about half of all
reported nearby SNe in the last half-decade were discovered by my team's Lick
Observatory Supernova Search (LOSS) conducted with the 0.76-m Katzman Automatic
Imaging Telescope (KAIT; Li et al. 2000; Filippenko et al. 2001; Filippenko
2003; see http://astro.berkeley.edu/$\sim$bait/kait.html). During LOSS, CCD
images are taken of $\simgt 1000$ galaxies per night and compared with KAIT
``template images'' obtained earlier; the templates are automatically
subtracted from the new images and analyzed with computer software, and the SN
candidates are flagged. The next day, undergraduate students at UC Berkeley
examine all candidates, including weak ones, to eliminate star-like cosmic
rays, asteroids, and other sources of false alarms. They also glance at some of
the subtracted images to locate SNe that might be near bright, poorly
subtracted stars or galactic nuclei. LOSS discovered 20 SNe in 1998, 40 in
1999, 38 in 2000, 68 in 2001, 82 in 2002, and 95 in 2003, making it by far the
world's leading search for nearby SNe.

   Spectroscopic classifications of these and other low-redshift SNe are
provided by a number of groups (see the {\it IAU Circulars}), including our own
observations with the Lick 3-m and Keck telescopes, the CfA SN monitoring
campaign, the Texas/McDonald Observatory group, the Asiago team, the European
Research and Training Network, the Australian National University team,
etc. The most important objects were photometrically monitored with KAIT
through \bvri (and sometimes $U$) filters (e.g., Li et al. 2001a, 2003b;
Modjaz et al. 2001; Leonard et al. 2002a,b; Foley et al. 2003), and unfiltered
follow-up observations (e.g., Matheson et al. 2001) were made of almost all of
them during the course of LOSS. Note that recently, KAIT has also been used to
automatically monitor the optical afterglows of gamma-ray bursts (Li et
al. 2003a,c; Matheson et al. 2003).

This growing sample of well-observed SNe~Ia allows us to more precisely
calibrate the distance determinations, as well as to look for correlations
between the observed properties of the SNe and their environment (Hubble type
of host galaxy, metallicity, stellar population, etc.). Jha (2002) present
\ubvri follow-up observations from the CfA SN monitoring campaign of 44 nearby
SNe~Ia (including many discovered by KAIT), from which they derive ``MLCS2k2,''
an updated light-curve fitting technique that includes $U$-band templates
(critical for SN~Ia observations at high redshift). Recently, progress has also
been made in the near-infrared, with Krisciunas et al. (2001, 2003, 2004)
presenting $JHK$ light curves of nearby SNe~Ia. These have enabled Krisciunas,
Phillips, \& Suntzeff (2004) to construct near-infrared Hubble diagrams of
SNe~Ia, with significantly less effect from dust extinction along the line of
sight. Their results suggest that SNe~Ia are much closer to {\it standard}
candles in the near-infrared than in the optical, with little dependence of
light-curve shape on near-infrared luminosity.

\section{Cosmological Uses: High Redshifts}

   These same techniques can be applied to construct a Hubble diagram with
high-redshift SNe~Ia, from which the value of $q_0 = (\Omega_M/2) -
\Omega_\Lambda$ can be determined. With enough objects spanning a range of
redshifts, we can measure $\Omega_M$ and $\Omega_\Lambda$ independently (e.g.,
Goobar \& Perlmutter 1995). Contours of peak apparent $R$-band magnitude for
SNe~Ia at two redshifts have different slopes in the
$\Omega_M$--$\Omega_\Lambda$ plane, and the regions of intersection provide the
answers we seek.

\subsection{The Search}

   Based on the pioneering work of Norgaard-Nielsen et al. (1989), whose goal
was to find SNe in moderate-redshift clusters of galaxies, the SCP (Perlmutter
et al. 1995a, 1997) and the HZT (Schmidt et al. 1998) devised a strategy that
almost guarantees the discovery of many faint, distant SNe~Ia ``on demand,''
during a predetermined set of nights.  This ``batch'' approach to studying
distant SNe allows follow-up spectroscopy and photometry to be scheduled in
advance, resulting in a systematic study not possible with random discoveries.
Most of the searched fields are equatorial, permitting follow-up from both
hemispheres.  The SCP was the first group to convincingly demonstrate the
ability to find SNe in batches.

    Our approach is simple in principle; see Schmidt et al. (1998) for details,
and for a description of our first high-redshift SN~Ia (SN 1995K). Pairs of
first-epoch images during the nights around new moon are obtained with
wide-field cameras on large telescopes (e.g., the Big Throughput Camera on the
CTIO 4-m Blanco telescope, and more recently the CTIO Mosaic II, the CFHT 8K
and 12K mosaics, and Suprime Cam on Subaru), followed by second-epoch images
3--4 weeks later.  (Pairs of images permit removal of cosmic rays, asteroids,
and Kuiper-belt objects.) These are compared immediately using well-tested
software, and new SN candidates are identified in the second-epoch images
(Figure 2). Spectra are obtained as soon as possible after discovery to verify
that the objects are SNe~Ia and to determine their redshifts.  Each team has
already found over 200 SNe in concentrated batches, as reported in numerous
{\it IAU Circulars} (e.g., Perlmutter et al. 1995b, 11 SNe with $0.16 \simlt z
\simlt 0.65$; Suntzeff et al. 1996, 17 SNe with $0.09
\simlt z \simlt 0.84$). 

\hbox{
\hskip 0.4truein
\vbox{\hsize 3.5 truein
\psfig{figure=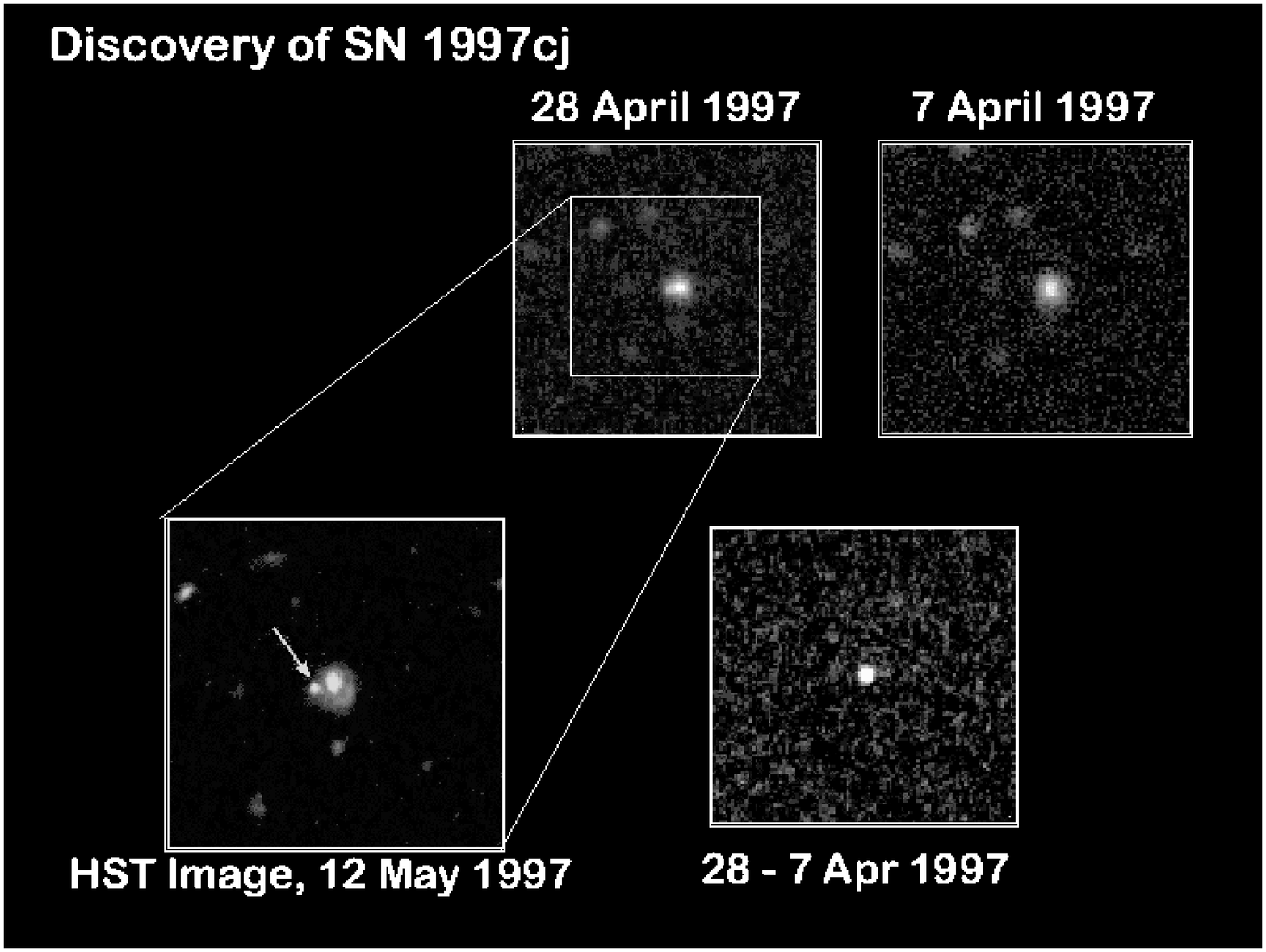,height=3.5truein,angle=270}
}
}

\smallskip
\noindent
{\it Figure 2:} Discovery image of SN 1997cj (28 April 1997), along with the
template image and an {\it HST} image obtained subsequently. The net
(subtracted) image is also shown.

\bigskip

   Intensive photometry of the SNe~Ia commences within a few days after
procurement of the second-epoch images; it is continued throughout the ensuing
and subsequent dark runs. In a few cases {\it HST} images are obtained. As
expected, most of the discoveries are {\it on the rise or near maximum
brightness}.  When possible, the SNe are observed in filters that closely match
the redshifted $B$ and $V$ bands; this way, the K-corrections become only a
second-order effect (Kim, Goobar, \& Perlmutter 1996; Nugent, Kim, \&
Perlmutter 2002).  Observations through two filters allow us to apply reddening
and luminosity corrections (Riess et al. 1996a; Hamuy et al. 1996a,b; Phillips
et al. 1999).

  Although SNe in the magnitude range 22--22.5 can sometimes be
spectroscopically confirmed with 4-m class telescopes, the signal-to-noise
ratios are low, even after several hours of integration. Certainly Keck,
Gemini, the VLT, Subaru, or Magellan are required for the fainter objects
(22.5--24.5 mag). With the largest telescopes, not only can we rapidly confirm
a substantial number of candidate SNe, but we can search for peculiarities in
the spectra that might indicate evolution of SNe~Ia with redshift.  Moreover,
high-quality spectra allow us to measure the age of a SN: we have developed a
method for automatically comparing the spectrum of a SN~Ia with a library of
spectra from many different epochs in the development of SNe~Ia (Riess et
al. 1997).  Our technique also has great practical utility at the telescope: we
can determine the age of a SN ``on the fly,'' within half an hour after
obtaining its spectrum. This allows us to decide rapidly which SNe are best for
subsequent photometric follow-up, and we immediately alert our collaborators
elsewhere.

\subsection{Results}

   First, we note that the light curves of high-redshift SNe~Ia are broader
than those of nearby SNe~Ia; the initial indications (Leibundgut et al. 1996;
Goldhaber et al. 1997), based on small numbers of SNe~Ia, are amply confirmed
with the larger samples (Goldhaber et al. 2001). Quantitatively, the amount by
which the light curves are ``stretched'' is consistent with a factor of $1 +
z$, as expected if redshifts are produced by the expansion of space rather than
by ``tired light'' and other non-expansion hypotheses for the redshifts of
objects at cosmological distances. [For non-standard cosmological
interpretations of the SN~Ia data, see Narlikar \& Arp (1997) and Hoyle,
Burbidge, \& Narlikar (2000).]  We also demonstrate this {\it
spectroscopically} at the $2\sigma$ confidence level for a single object: the
spectrum of SN 1996bj ($z = 0.57$) evolved more slowly than those of nearby
SNe~Ia, by a factor consistent with $1 + z$ (Riess et al. 1997).  Observations
of SN 1997ex ($z = 0.36$) at three epochs conclusively verify the effects of
time dilation: temporal changes in the spectra are slower than those of nearby
SNe~Ia by roughly the expected factor of 1.36. Although one might be able to
argue that something other than universal expansion could be the cause of the
apparent stretching of SN~Ia light curves at high redshifts, it is much more
difficult to attribute apparently slower evolution of spectral details to an
unknown effect.

   The formal value of $\Omega_M$ derived from SNe~Ia has changed with time.
The SCP published the initial result (Perlmutter et al. 1995a), based on a
single object, SN 1992bi at $z = 0.458$: $\Omega_M = 0.2 \pm 0.6 \pm 1.1$
(assuming that $\Omega_\Lambda = 0$). The SCP's analysis of their first seven
objects (Perlmutter et al. 1997) suggested a much larger value of $\Omega_M =
0.88 \pm 0.6$ (if $\Omega_\Lambda = 0$) or $\Omega_M = 0.94 \pm 0.3$ (if
$\Omega_{\rm total} = 1$). Such a high-density universe seemed at odds with
other, independent measurements of $\Omega_M$ at that time. However, with the
subsequent inclusion of just one more object, SN 1997ap at $z = 0.83$ (the
highest then known for a SN~Ia; Perlmutter et al. 1998), their estimates were
revised back down to $\Omega_M = 0.2 \pm 0.4$ if $\Omega_\Lambda = 0$, and
$\Omega_M = 0.6 \pm 0.2$ if $\Omega_{\rm total} = 1$; the apparent brightness
of SN 1997ap had been precisely measured with {\it HST}, so it substantially
affected the best fits.

   Meanwhile, the HZT published (Garnavich et al. 1998a) an analysis of four
objects (three of them observed with {\it HST}), including SN 1997ck at $z =
0.97$, at that time a redshift record, although they cannot be absolutely
certain that the object was a SN~Ia because the spectrum is too poor. From
these data, the HZT derived that $\Omega_M = -0.1 \pm 0.5$ (assuming
$\Omega_\Lambda = 0$) and $\Omega_M = 0.35 \pm 0.3$ (assuming $\Omega_{\rm
total} = 1$), inconsistent with the high $\Omega_M$ initially found by
Perlmutter et al. (1997) but consistent with the revised estimate in Perlmutter
et al. (1998). An independent analysis of 10 SNe~Ia using the ``snapshot''
distance method (with which conclusions are drawn from sparsely observed
SNe~Ia) gave quantitatively similar conclusions (Riess et al. 1998a). However,
none of these early data sets carried the statistical discriminating power to
detect cosmic acceleration.

   The SCP's next results were announced at the 1998 January AAS meeting in
Washington, DC. A press conference was scheduled, with the stated purpose of
presenting and discussing the then-current evidence for a low-$\Omega_M$
universe as published by Perlmutter et al. (1998; SCP) and Garnavich et al.
(1998a; HZT). When showing the SCP's Hubble diagram for SNe~Ia, however,
Perlmutter also pointed out tentative evidence for {\it acceleration}! He
said that the conclusion was uncertain, and that the data were equally
consistent with no acceleration; the systematic errors had not yet been
adequately assessed. Essentially the same conclusion was given by the SCP
in February 1998 during their talks at a major conference on dark matter
held near Los Angeles (Goldhaber \& Perlmutter 1998).

   Although it chose not to reveal them at the same 1998 January AAS meeting,
the HZT already had similar, tentative evidence for acceleration in their own
SN~Ia data set. The HZT continued to perform numerous checks of their data
analysis and interpretation, including fairly thorough consideration of various
possible systematic effects. Unable to find any significant problems, even with
the possible systematic effects, the HZT reported detection of a {\it nonzero}
value for $\Omega_\Lambda$ (based on 16 high-$z$ SNe~Ia) at the Los Angeles
dark matter conference in February 1998 (Filippenko \& Riess 1998), and soon
thereafter submitted a formal paper that was published in September 1998 (Riess
et al. 1998b). 

\bigskip

\hbox{
\hskip -0.15truein
\vbox{\hsize 2.2 truein
\psfig{figure=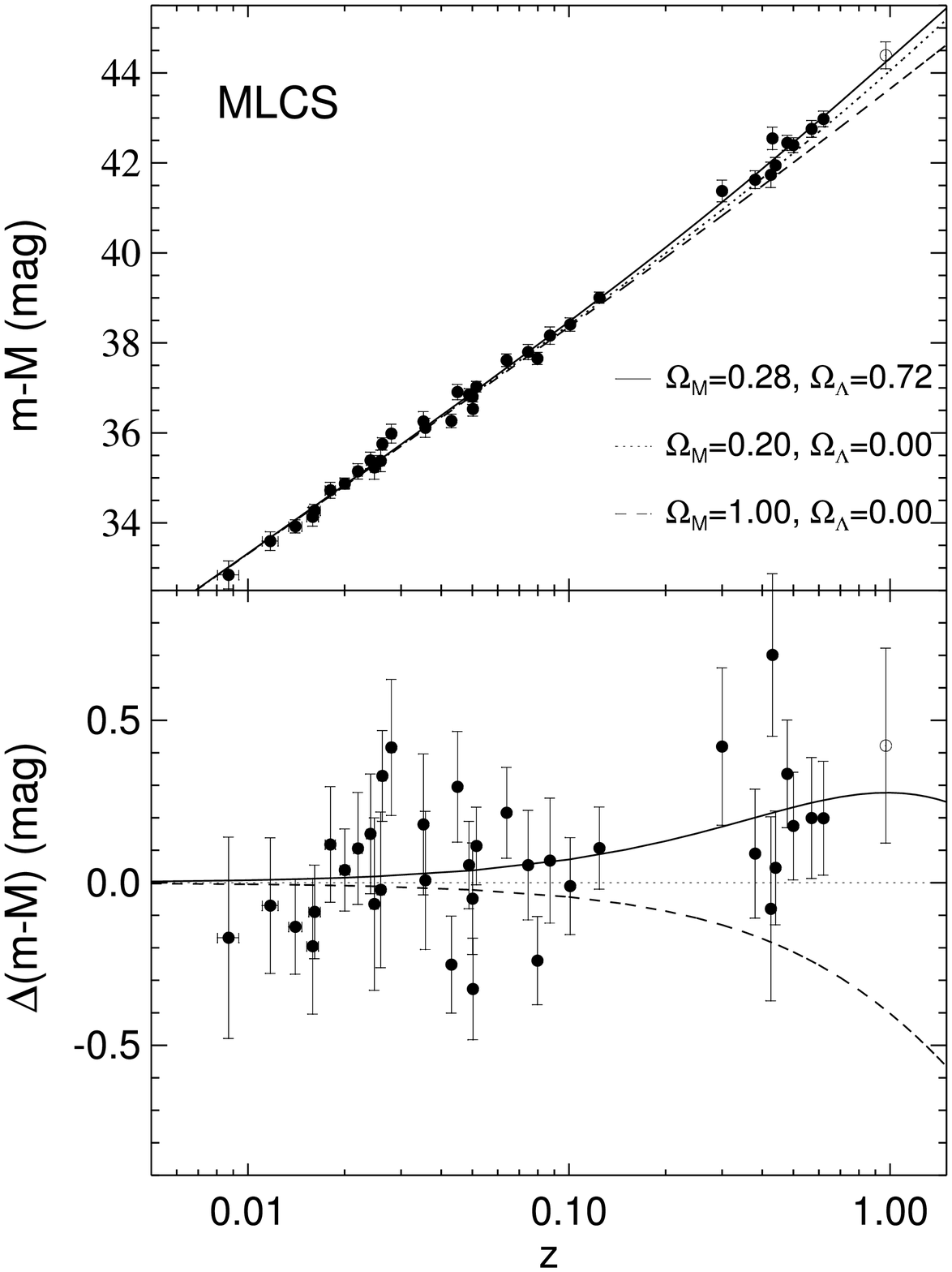,height=2.9truein,angle=0}
}
\hskip +0.0truein
\vbox{\hsize 3.0 truein
\psfig{figure=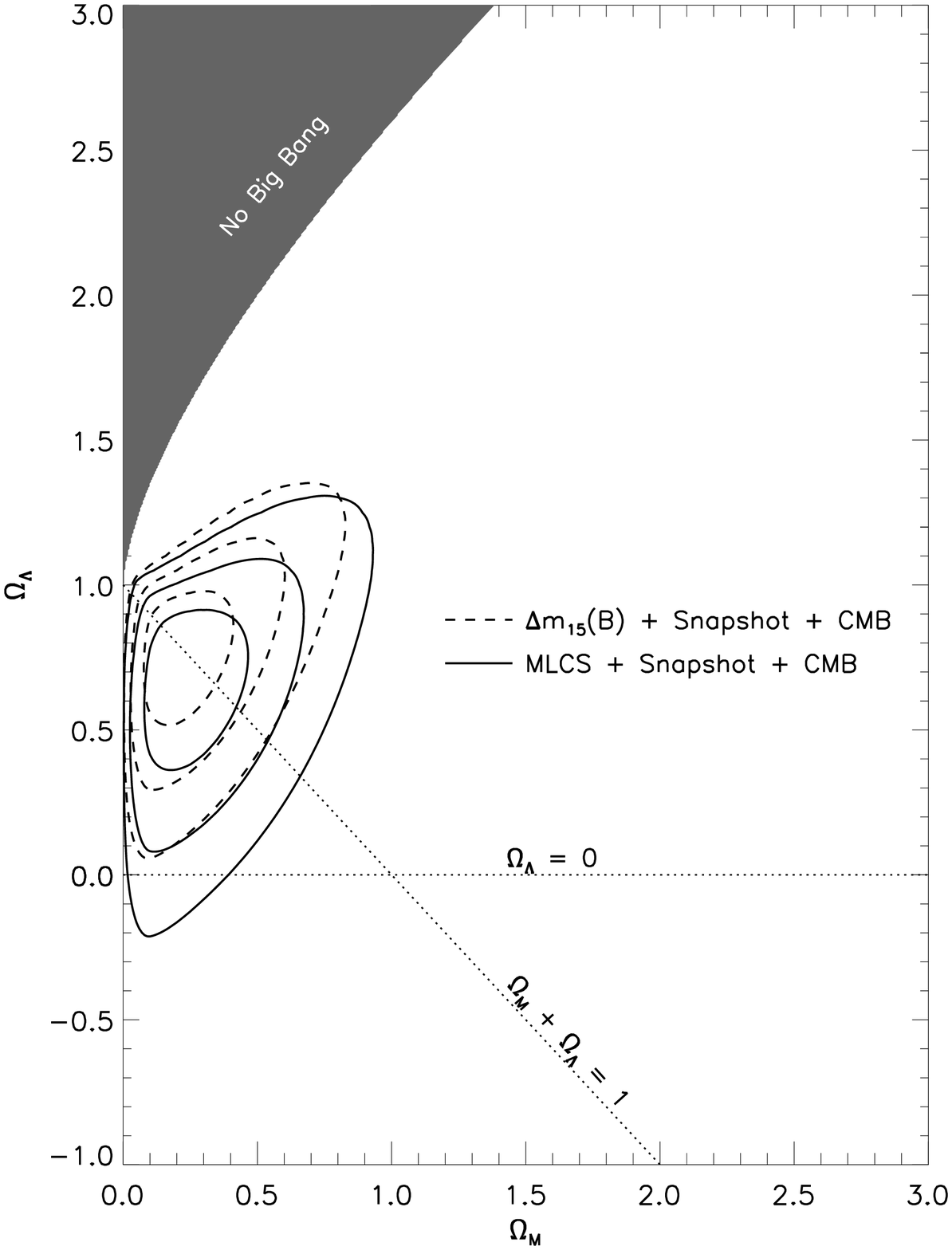,height=2.9truein,angle=0}
}
}

\noindent
{\it Figure 3 (left):} The upper panel shows the Hubble diagram for the low-$z$
and high-$z$ HZT SN~Ia sample with MLCS distances; see Riess et
al. (1998b). Overplotted are three world models: ``low'' and ``high''
$\Omega_M$ with $\Omega_\Lambda=0$, and the best fit for a flat universe
($\Omega_M = 0.28$, $\Omega_\Lambda = 0.72$).  The bottom panel shows the
difference between data and models from the $\Omega_M=0.20$, $\Omega_\Lambda=0$
prediction.  Only the 10 best-observed high-$z$ SNe~Ia are shown.  The average
difference between the data and the $\Omega_M=0.20$, $\Omega_\Lambda=0$
prediction is $\sim 0.25$ mag.

\medskip
\noindent
{\it Figure 4 (right):} The HZT's combined constraints from SNe~Ia (left) and
the position of the first acoustic peak of the cosmic microwave background
(CMB) angular power spectrum, based on data available in mid-1998; see
Garnavich et al. (1998b).  The contours mark the 68.3\%, 95.4\%, and 99.7\%
enclosed probability regions. Solid curves correspond to results from the MLCS
method; dotted ones are from the $\Delta m_{15}(B)$ method; all 16 SNe~Ia in
Riess et al. (1998b) were used.

\bigskip

  The HZT's original Hubble diagram for the 10 best-observed high-$z$
SNe~Ia is given in Figure 3. With the MLCS method applied to the 
full set of 16 SNe~Ia, the HZT's formal results were $\Omega_M = 0.24 \pm 
0.10$ if $\Omega_{\rm total} = 1$, or $\Omega_M = -0.35 \pm 0.18$ (unphysical) 
if $\Omega_\Lambda = 0$. If one demanded that $\Omega_M = 0.2$, then the best
value for $\Omega_\Lambda$ was $0.66 \pm 0.21$.  These conclusions did not
change significantly when only the 10 best-observed SNe~Ia were used 
(Figure 3; $\Omega_M = 0.28 \pm 0.10$ if $\Omega_{\rm total} = 1$).

  Another important constraint on the cosmological parameters could be obtained
from measurements of the angular scale of the first acoustic peak of the CMB
(e.g., Zaldarriaga, Spergel, \& Seljak 1997; Eisenstein, Hu, \& Tegmark 1998);
the SN~Ia and CMB techniques provide nearly complementary information. A
stunning result was already available by mid-1998 from existing measurements
(e.g., Hancock et al. 1998; Lineweaver \& Barbosa 1998): the HZT's analysis of
the SN~Ia data in Riess et al. (1998b) demonstrated that $\Omega_M +
\Omega_\Lambda = 0.94 \pm 0.26$ (Figure 4), when the SN and CMB constraints
were combined (Garnavich et al. 1998b; see also Lineweaver 1998, Efstathiou et
al. 1999, and others).

   Somewhat later (June 1999), the SCP published almost identical results
(Perlmutter et al. 1999), implying an accelerating expansion of the Universe,
based on an essentially independent set of 42 high-$z$ SNe (although a
substantial number of these were later discarded because they are not clearly
SNe~Ia; Knop et al.  2003). The Perlmutter et al. (1999) data, together with
those of the HZT, are shown in Figure 5, and the corresponding confidence
contours in the $\Omega_\Lambda$ vs.  $\Omega_M$ plane are given in Figure
6. This incredible agreement suggested that neither group had made a large,
simple blunder; if the result was wrong, the reason must be subtle.  Had there
been only one team working in this area, it is likely that far fewer
astronomers and physicists throughout the world would have taken the result
seriously.

   Moreover, already in 1998--1999 there was tentative evidence that the ``dark
energy'' driving the accelerated expansion was indeed consistent with the
cosmological constant, $\Lambda$. If $\Lambda$ dominates, then the
equation-of-state parameter of the dark energy should be $w = -1$, where the
pressure ($P$) and density ($\rho$) are related according to $w = P/(\rho
c^2)$.  Garnavich et al. (1998b) and Perlmutter et al. (1999) were able to set
an interesting limit, $w \simlt -0.60$ at the 95\% confidence level. However,
more high-quality data at $z \approx 0.5$ are needed to narrow the allowed
range, in order to test other proposed candidates for dark energy such as
various forms of ``quintessence'' (e.g., Caldwell, Dav\'e, \& Steinhardt 1998).

   Although the CMB results appeared reasonably persuasive in 1998--1999, one
could argue that fluctuations on different scales had been measured with
different instruments, and that subtle systematic effects might lead to
erroneous conclusions. These fears were dispelled only 1--2 years later, when
the more accurate and precise results of the BOOMERANG collaboration were
announced (de Bernardis et al. 2000, 2002). Shortly thereafter the MAXIMA
collaboration distributed their very similar findings (Hanany et al. 2000;
Balbi et al. 2000; Netterfield et al. 2002; see also the TOCO, DASI, and many
other measurements).  Figure 6 illustrates that the CMB measurements tightly
constrain $\Omega_{\rm total}$ to be close to unity; we appear to live in a
flat universe, in agreement with most inflationary models for the early
Universe! Combined with the SN~Ia results, the evidence for nonzero
$\Omega_\Lambda$ was fairly strong. Making the argument even more compelling
was the fact that various studies of clusters of galaxies (see summary by
Bahcall et al. 1999) showed that $\Omega_M \approx 0.3$, consistent with the
results in Figures 4 and 6. Thus, a ``concordance cosmology'' had emerged:
$\Omega_M \approx 0.3$, $\Omega_\Lambda \approx 0.7$ --- consistent with what
had been suspected some years earlier by Ostriker \& Steinhardt (1995; see also
Carroll, Press, \& Turner 1992).

\medskip

\hbox{
\hskip -0.25truein
\vbox{\hsize 2.2 truein
\psfig{figure=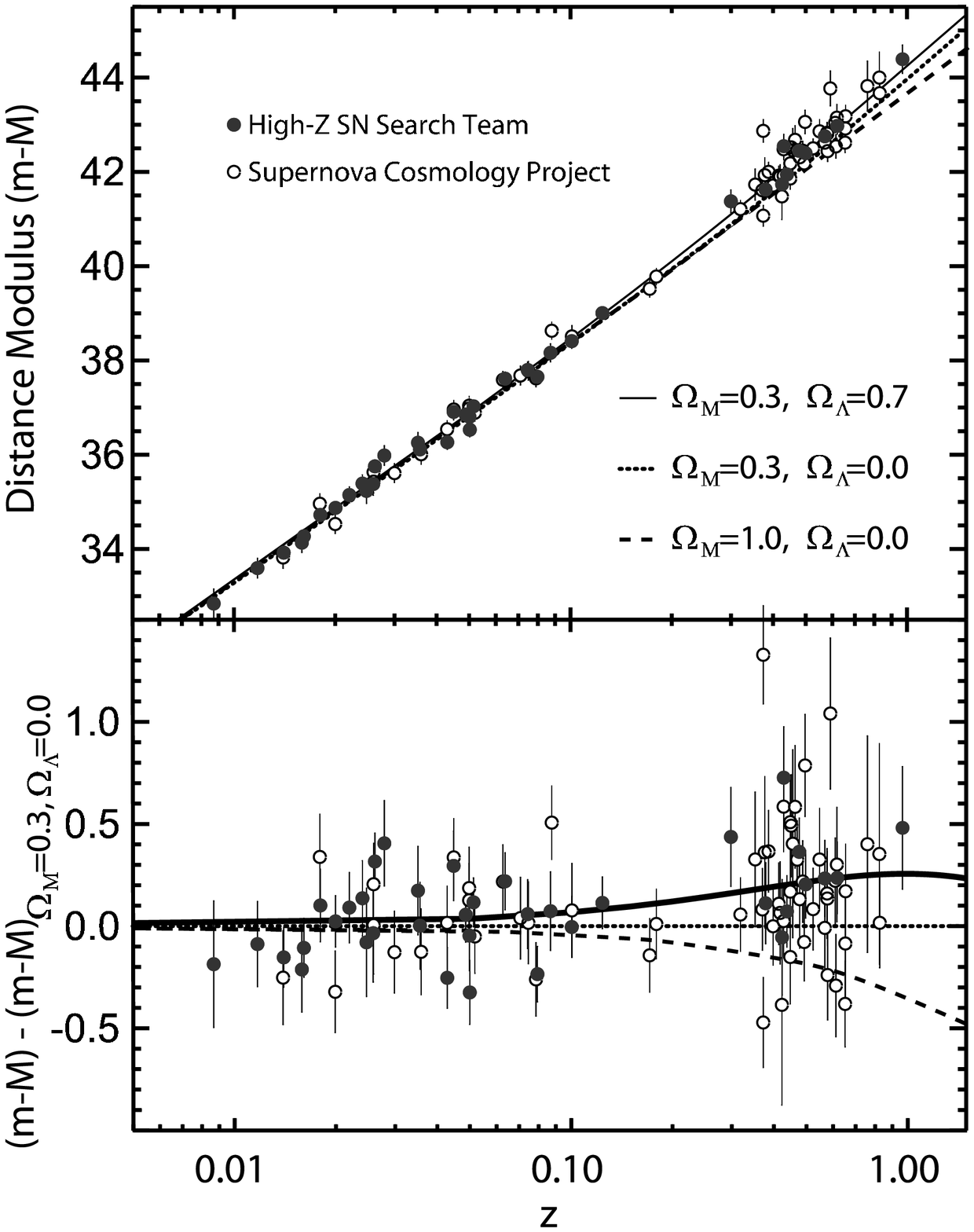,height=2.9truein,angle=0}
}
\hskip 0.1truein
\vbox{\hsize 2.5 truein
\psfig{figure=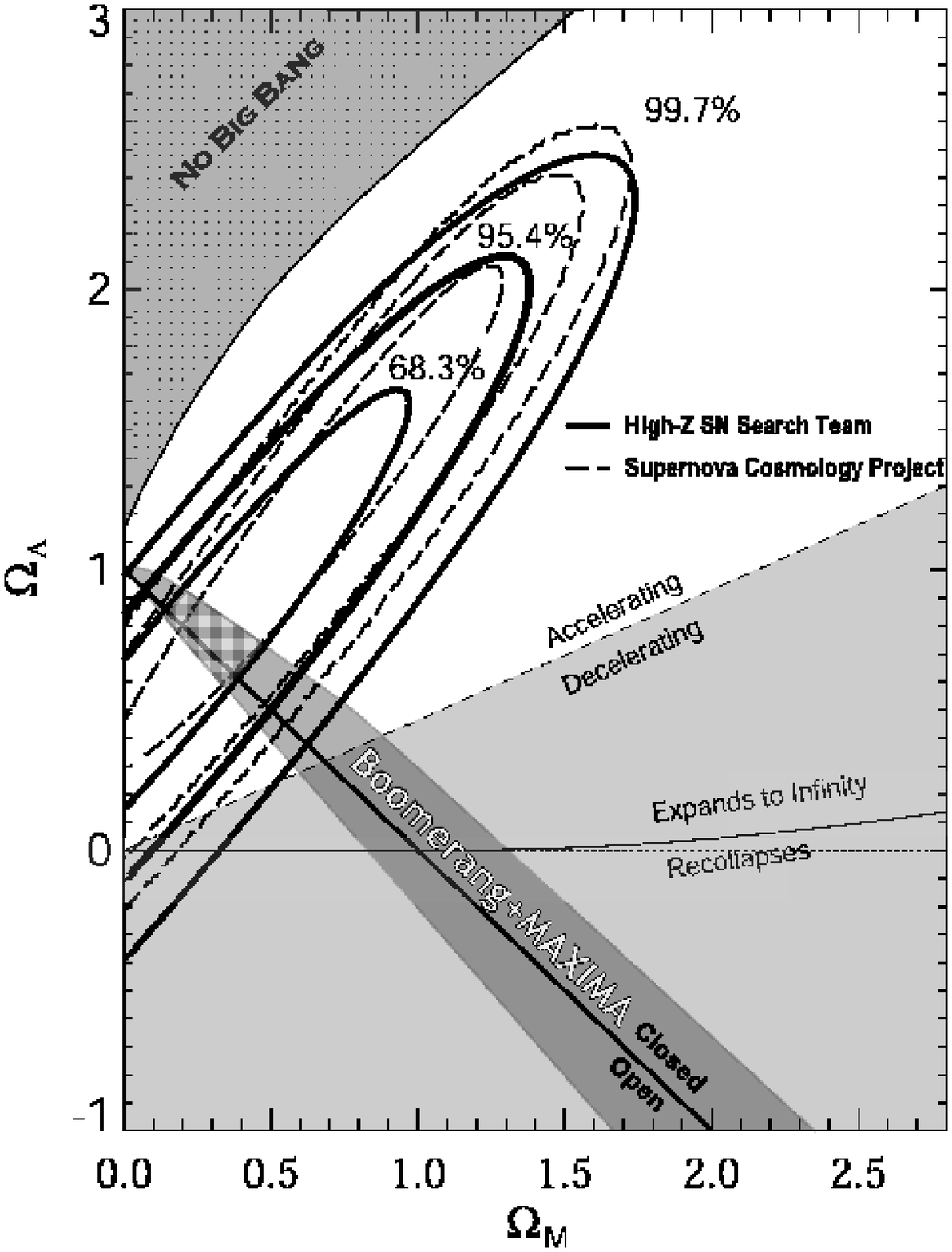,height=2.9truein,angle=0}
}
}

\vskip -0.3truein
\noindent
{\it Figure 5 (left):} As in Figure 3, but this time including both the HZT
(Riess et al. 1998b) and SCP (Perlmutter et al. 1999) samples of low-redshift
and high-redshift SNe~Ia. Overplotted are three world models: $\Omega_M = 0.3$
and $1.0$ with $\Omega_\Lambda=0$, and a flat universe ($\Omega_{\rm total} =
1.0$) with $\Omega_\Lambda = 0.7$.  The bottom panel shows the difference
between data and models from the $\Omega_M=0.3$, $\Omega_\Lambda=0$ prediction.

\smallskip
\noindent
{\it Figure 6 (right):} The combined constraints from SNe~Ia (see Figure 5)
and the position of the first acoustic peak of the CMB angular power spectrum,
based on BOOMERANG and MAXIMA data.  The contours mark the 68.3\%, 95.4\%, and
99.7\% enclosed probability regions determined from the SNe~Ia. According to
the CMB, $\Omega_{\rm total} \approx 1.0$.

\bigskip

  Yet another piece of evidence for a nonzero value of $\Lambda$ was provided
by the Two-Degree Field Galaxy Redshift Survey (2dFGRS; Peacock et al. 2001;
Percival et al. 2001; Efstathiou et al. 2002). Combined with the CMB maps,
their results are inconsistent with a universe dominated by gravitating dark
matter. Again, the implication is that about 70\% of the mass-energy density of
the Universe consists of some sort of dark energy whose gravitational effect is
repulsive. In 2003, results from the first year of {\it Wilkinson Microwave
Anisotropy Probe (WMAP)}\ observations appeared; together with the 2dFGRS
constraints, they confirmed and refined the concordance cosmology ($\Omega_M =
0.27$, $\Omega_\Lambda = 0.73$, $\Omega_{\rm baryon} = 0.044$, $H_0 = 71 \pm 4$
km s$^{-1}$ Mpc$^{-1}$; Spergel et al. 2003, and references therein).  Recent
evidence for dark energy and an accelerating universe has also come from
2--3$\sigma$ detections of the integrated Sachs-Wolfe effect (Afshordi, Loh,
\& Strauss 2004; Boughn \& Crittenden 2004; Fosalba et al. 2003; Nolta et
al. 2004; Scranton et al. 2004).

   Meanwhile, the SN~Ia measurements were becoming more numerous and of higher
quality. For the HZT, the new (Fall 1999) sample of high-redshift SNe~Ia
presented by Tonry et al. (2003), analyzed with methods distinct from (but
similar to) those used previously, confirmed the result of Riess et al. (1998b)
and Perlmutter et al. (1999) that the expansion of the Universe is
accelerating. By combining {\it all} of the data sets available at that time,
Tonry et al. (2003) were able to use about 200 SNe~Ia, obtaining an incredibly
firm detection of $\Omega_\Lambda > 0$. They placed the following constraints
on cosmological quantities: (1) If the equation-of-state parameter of the dark
energy is $w = -1$, then $H_0 t_0 = 0.96 \pm 0.04$, and $\Omega_\Lambda - 1.4
\Omega_M = 0.35 \pm 0.14.$ (2) Including the constraint of a flat universe,
they find that $\Omega_M = 0.28 \pm 0.05$, independent of any large-scale
structure measurements. (3) Adopting a prior based on the 2dFGRS constraint on
$\Omega_M$ (Percival et al. 2001) and assuming a flat universe, they derive
that $-1.48 < w < -0.72$ at 95\% confidence. (4) Adopting the 2dFGRS results,
they find $\Omega_M = 0.28$ and $\Omega_\Lambda = 0.72$, independent of any
assumptions about $\Omega_{\rm total}$.  These constraints are similar in
precision and in value to conclusions reported using Year 1 of {\it WMAP}\
(Spergel et al. 2003), also in combination with the 2dFGRS. 

   Even more recently, Barris et al. (2004) describe results from the 2001 HZT
campaign. 23 SNe with $z = 0.34$--1.03 were studied, 9 of which are
unambiguously SNe~Ia. The sample includes 15 objects at $z > 0.7$, doubling
the number of published SNe at these redshifts. Under the assumption that
$\Omega_{\rm total} = 1$, they determine best-fit values of ($\Omega_M, \,
\Omega_\Lambda$) = (0.33, 0.67), using 22 SNe from the 2001 HZT survey 
together with 172 SNe from Tonry et al. (2003) satisfying $z > 0.01$ and
$A_V \leq 0.5$ mag.

   The SCP's high-quality {\it HST} data set on 11 SNe~Ia in the redshift range
0.36--0.86, recently published by Knop et al. (2003), independently confirms
the apparent acceleration of the Universe. They were able to measure accurate
colors of the SNe, providing better host-galaxy extinction estimates than had
been possible in the past for individual objects. Thus, there was no need to
make any assumptions or priors on the parent distribution of extinction values,
$E(B-V)$, unlike the case in the initial analyses (Riess et al.  1998;
Perlmutter et al. 1999; but see Sullivan et al. 2003, discussed below).  Their
extinction measurements do not show evidence for anomalously blue SNe~Ia at
high redshifts (or for a preponderance of negative $E(B-V)$ values for
high-redshift SNe~Ia), contrary to some earlier suspicions (Falco et al. 1999;
Leibundgut 2001).  Knop et al. (2003) find that dark energy is required with a
probability exceeding 99\%, consistent with previous studies that had made
assumptions about the distribution of extinctions or that had used
low-extinction subsets of SNe~Ia. In a flat universe with a constant dark
energy equation-of-state parameter $w = -1$, they find that $\Omega_\Lambda =
0.75~ (+0.06, -0.07) \pm 0.04$, where the first quoted uncertainties are
statistical and the second are identified systematics. Moreover, in a flat
universe with $w$ constant in time, their SNe~Ia data show that $w = -1.05
(+0.15, -0.20) \pm 0.09$, consistent with $\Lambda$ (rather than with some
quintessence models; see below).

   The dynamical age of the Universe can be calculated from the cosmological
parameters. In an empty Universe with no cosmological constant, the dynamical
age is simply the inverse of the Hubble constant, $t_0 = H_0^{-1}$; there is no
deceleration. In the late-1990s, SNe~Ia gave $H_0 = 65
\pm 7$ km s$^{-1}$ Mpc$^{-1}$, and a Hubble time of $15.1 \pm 1.6$ Gyr. For a
more complex cosmology, integrating the velocity of the expansion from the
current epoch ($z=0$) to the beginning ($z=\infty$) yields an expression for
the dynamical age. As shown in detail by Riess et al. (1998b), by mid-1998 the
HZT had obtained a value of 14.2$^{+1.0}_{-0.8}$ Gyr (with $H_0 = 65$) using
the likely range for $(\Omega_M, \Omega_\Lambda)$ that they measured.  (The
precision was so high because their experiment was sensitive to roughly the
{\it difference} between $\Omega_M$ and $\Omega_\Lambda$, and the dynamical age
also varies in approximately this way.)  Including the {\it systematic}
uncertainty of the Cepheid distance scale, which may be up to 10\%, a
reasonable estimate of the dynamical age was $14.2 \pm 1.7$ Gyr (Riess et
al. 1998b). Again, the SCP's result was very similar (Perlmutter et al. 1999),
since it was based on nearly the same derived values for the cosmological
parameters. The most recent results, reported by Tonry et al. (2003; see also
Knop et al. 2003) and adopting $H_0 = 72 \pm 8$ km s$^{-1}$ Mpc$^{-1}$, give a
dynamical age of $13.1 \pm 1.5$ Gyr for the Universe --- again, in agreement
with the {\it WMAP} result of $13.7 \pm 0.2$ Gyr.

   This expansion age is also consistent with ages determined from various
other techniques such as the cooling of white dwarfs (Galactic disk $> 9.5$
Gyr; Oswalt et al. 1996), radioactive dating of stars via the thorium and
europium abundances ($15.2 \pm 3.7$ Gyr; Cowan et al.  1997), and studies of
globular clusters (10--15 Gyr, depending on whether {\it Hipparcos} parallaxes
of Cepheids are adopted; Gratton et al. 1997; Chaboyer et al. 1998).  The ages
of the oldest stars no longer seem to exceed the expansion age of the Universe;
the long-standing ``age crisis'' has evidently been resolved.

\section{Discussion}

   Although the convergence of different methods on the same answer
is reassuring, and suggests that the concordance cosmology is correct, 
it is important to vigorously test each method to make sure it is not
leading us astray. Moreover, only through such detailed studies will
the accuracy and precision of the methods improve, allowing us to eventually
set better constraints on the equation-of-state parameter, $w$. Here I
discuss the systematic effects that could adversely affect the SN~Ia results.

   High-redshift SNe~Ia are observed to be dimmer than expected in an empty
Universe (i.e., $\Omega_M = 0$) with no cosmological constant. At $z \approx
0.5$, where the SN~Ia observations have their greatest leverage on $\Lambda$,
the difference between an $\Omega_M = 0.3$
($\Omega_\Lambda = 0$) universe and a flat universe with $\Omega_\Lambda =0.7$
is only about 0.25 mag. Thus, we need to find out if chemical abundances,
stellar populations, selection bias, gravitational lensing, or grey dust can
have an effect this large. Although both the HZT and SCP had considered many of
these potential systematic effects in their original discovery papers (Riess et
al. 1998b; Perlmutter et al. 1999), and had shown with reasonable confidence
that obvious ones were not greatly affecting their conclusions, it was of
course possible that they were wrong, and that the data were being
misinterpreted.

\subsection{Gravitational Lensing}

  The magnification and demagnification of light by large-scale structure can
alter the observed magnitudes of high-redshift SNe (e.g., Kantowski, Vaughan,
\& Branch 1995; Kantowski 1998).  The effect of weak gravitational lensing on
our analysis has been quantified by Wambsganss, Cen, \& Ostriker (1998) and
summarized by Schmidt et al. (1998).  SN~Ia light will, on average, be
demagnified by 0.5\% at $z = 0.5$ and by 1\% at $z = 1$ in a Universe with a
non-negligible cosmological constant.  Although the sign of the effect is the
same as the influence of a cosmological constant, the size of the effect is
small enough to be ignored.

   Holz (1998), Holz \& Wald (1998), and Kantowski (1998) have calculated the
weak lensing effects on supernova light from ordinary matter which is not
smoothly distributed in galaxies but rather clumped into stars (i.e., dark
matter contained in massive compact halo objects).  With this scenario,
microlensing becomes a more important effect, further decreasing the typical
observed supernova luminosities at $z = 0.5$ by 0.02 mag for $\Omega_M$ = 0.2.
Even if most ordinary matter were contained in compact objects, this effect
would not be large enough to reconcile the SN~Ia distances with the influence
of ordinary matter alone in a decelerating universe. Barris et al. (2004), for
example, show that use of the ``empty-beam model'' for gravitational lensing
does not eliminate the need for $\Omega_\Lambda > 0$.  However, gravitational
lensing will certainly need to be taken into account when making a precise
measurement of $w$.

   With a very large sample (200--1000) of high-redshift SNe~Ia, and accurate
photometry, it should be possible to quantify the effects of clumped matter.
Light from some SNe~Ia should be strongly amplified by the presence of
intervening matter, while the vast majority will be deamplified (e.g., Barber
et al.  2000). The distribution of amplification factors can be used to
determine the type of dark matter most prevalent in the Universe (compact
objects, or smoothly distributed).

\subsection{Evolution}

   Perhaps the most obvious possible systematic effect is {\it evolution} of
SNe~Ia over cosmic time, due to changes in metallicity, progenitor mass, or
some other factor. If the peak luminosity of SNe~Ia were lower at high
redshift, then the case for $\Omega_\Lambda > 0$ would weaken.  Conversely, if
the distant explosions are more powerful, then the case for acceleration
strengthens.  Theorists are not yet sure what the sign of the effect will be,
if it is present at all; different assumptions lead to different conclusions
(H\"{o}flich, Wheeler, \& Thielemann 1998; Umeda et al. 1999; Nomoto et
al. 2000; Yungelson \& Livio 2000).

     Of course, it is extremely difficult, if not effectively impossible, to
obtain an accurate, independent measure of the peak luminosity of high-$z$
SNe~Ia, and hence to directly test for luminosity evolution. However, we can
more easily determine whether {\it other} observable properties of low-$z$ and
high-$z$ SNe~Ia differ. If they are all the same, it is more probable that the
peak luminosity is constant as well --- but if they differ, then the peak
luminosity might also be affected (e.g., H\"{o}flich et al. 1998).  Drell,
Loredo, \& Wasserman (2000), for example, argue that there are reasons to
suspect evolution, because the average properties of existing samples of
high-$z$ and low-$z$ SNe~Ia seem to differ (e.g., the high-$z$ SNe~Ia are more
uniform).

   The local sample of SNe~Ia displays a weak correlation between light-curve
shape and host-galaxy type, in the sense that the most luminous SNe~Ia with the
broadest light curves only occur in late-type galaxies.  Both early-type and
late-type galaxies provide hosts for dimmer SNe~Ia with narrower light curves
(Hamuy et al.  1996a). The mean luminosity difference for SNe~Ia in late-type
and early-type galaxies is $\sim 0.3$ mag.  In addition, the SN~Ia rate per
unit luminosity is almost twice as high in late-type galaxies as in early-type
galaxies at the present epoch (Cappellaro et al. 1997). These results may
indicate an evolution of SNe~Ia with progenitor age. Possibly relevant physical
parameters are the mass, metallicity, and C/O ratio of the progenitor
(H\"{o}flich et al. 1998).

   We expect that the relation between light-curve shape and peak luminosity
that applies to the range of stellar populations and progenitor ages
encountered in the late-type and early-type hosts in our nearby sample should
also be applicable to the range encountered in our distant sample.  In fact,
the age range for SN~Ia progenitors in the nearby sample is likely {\it larger}
than the change in mean progenitor age over the 4--6 Gyr lookback time to the
high-$z$ sample.  Thus, to first order at least, our local sample should
correct the distances for progenitor or age effects.

   We can place empirical constraints on the effect that a change in the
progenitor age would have on our SN~Ia distances by comparing subsamples of
low-redshift SNe~Ia believed to arise from old and young progenitors.  In the
nearby sample, the mean difference between the distances for the early-type
hosts (8 SNe~Ia) and late-type hosts (19 SNe~Ia), at a given redshift, is 0.04
$\pm$ 0.07 mag from the MLCS method.  This difference is consistent with zero.
Even if the SN~Ia progenitors evolved from one population at low redshift to
the other at high redshift, we still would not explain the surplus in mean
distance of 0.25 mag over the $\Omega_\Lambda=0$ prediction. 

   Moreover, in a major study of high-redshift SNe~Ia as a function of galaxy
morphology, the SCP found no clear differences (except for the amount of
scatter; see \S 5.2) between the cosmological results obtained with SNe~Ia in
late-type and early-type galaxies (Sullivan et al. 2003). Similarly, a study of
the host galaxies of high-redshift SNe~Ia ($0.42 < z < 1.06$) done by the HZT
(Williams et al. 2003) found no clear evidence for correlations between
host-galaxy properties and the {\it residuals} of distance measurements from
cosmological fits. Some of the correlations between SN~Ia properties and
host-galaxy type seen at low redshift also appear to be present at high
redshift, again supporting the current practice of extrapolating properties of
the nearby SN~Ia population to high redshifts.

  It is also reassuring that initial comparisons suggest that high-$z$ SN~Ia
spectra appear similar to those observed at low redshift.  For example, the
spectral characteristics of SN 1998ai ($z = 0.49$) appear to be essentially
indistinguishable from those of normal low-$z$ SNe~Ia; see Figure 7. In fact,
the most obviously discrepant spectrum in this figure is the second one, that
of SN 1994B ($z = 0.09$); it is intentionally included as a ``decoy'' that
illustrates the degree to which even the spectra of nearby, relatively normal
SNe~Ia can vary. Nevertheless, it is important to note that a dispersion in
luminosity (perhaps 0.2 mag) exists even among the other, more normal SNe~Ia
shown in Figure 7; thus, our spectra of SN 1998ai and other high-$z$ SNe~Ia are
not yet sufficiently good for independent, {\it precise} determinations of peak
luminosity from spectral features (Nugent et al. 1995).  Many of them, however,
are sufficient for ruling out other SN types (Figure 8), or for identifying
gross peculiarities such as those shown by SNe 1991T and 1991bg; see Coil et
al. (2000).

  We can help verify that the SNe at $z \approx 0.5$ being used for cosmology
do not belong to a subluminous population of SNe~Ia by examining rest-frame
$I$-band light curves. Normal, nearby SNe~Ia show a pronounced second maximum
in the $I$ band about a month after the first maximum and typically about 0.5
mag fainter (e.g., Ford et al. 1993; Suntzeff 1996). Subluminous SNe~Ia, in
contrast, do not show this second maximum, but rather follow a linear decline
or show a muted second maximum (Filippenko et al. 1992a). As discussed by Riess
et al. (2000), tentative evidence for the second maximum is seen from the HZT's
existing $J$-band (rest-frame $I$-band) data on SN 1999Q ($z = 0.46$); see
Figure 10. Additional tests with spectra and near-infrared light curves are
currently being conducted.

  Another way of using light curves to test for possible evolution of SNe~Ia is
to see whether the rise time (from explosion to maximum brightness) is the same
for high-redshift and low-redshift SNe~Ia; a difference might indicate that the
peak luminosities are also different (H\"{o}flich et al. 1998). Riess et al.
(1999c) measured the risetime of nearby SNe~Ia, using data from KAIT, the
Beijing Astronomical Observatory (BAO) SN search, and a few amateur
astronomers. Though the exact value of the risetime is a function of peak
luminosity, for typical low-redshift SNe~Ia it is $20.0 \pm 0.2$ days. Riess et
al. (1999b) pointed out that this differs by $5.8\sigma$ from the {\it
preliminary} risetime of $17.5 \pm 0.4$ days reported in conferences by the SCP
(Goldhaber et al. 1998a,b; Groom 1998). However, more thorough analyses of the
SCP data (Aldering, Knop, \& Nugent 2000; Goldhaber et al. 2001) show that the
high-redshift uncertainty of $\pm 0.4$ days that the SCP originally reported
was much too small because it did not account for systematic effects. The
revised discrepancy with the low-redshift risetime is about $2\sigma$ or
less. Thus, the apparent difference in risetimes might be insignificant. Even
if the difference is real, however, its relevance to the peak luminosity is
unclear; the light curves may differ only in the first few days after the
explosion, and this could be caused by small variations in conditions near the
outer part of the exploding white dwarf that are inconsequential at the peak.

\medskip

\hbox{
\hskip -0.25truein
\vbox{\hsize 2.15 truein
\psfig{figure=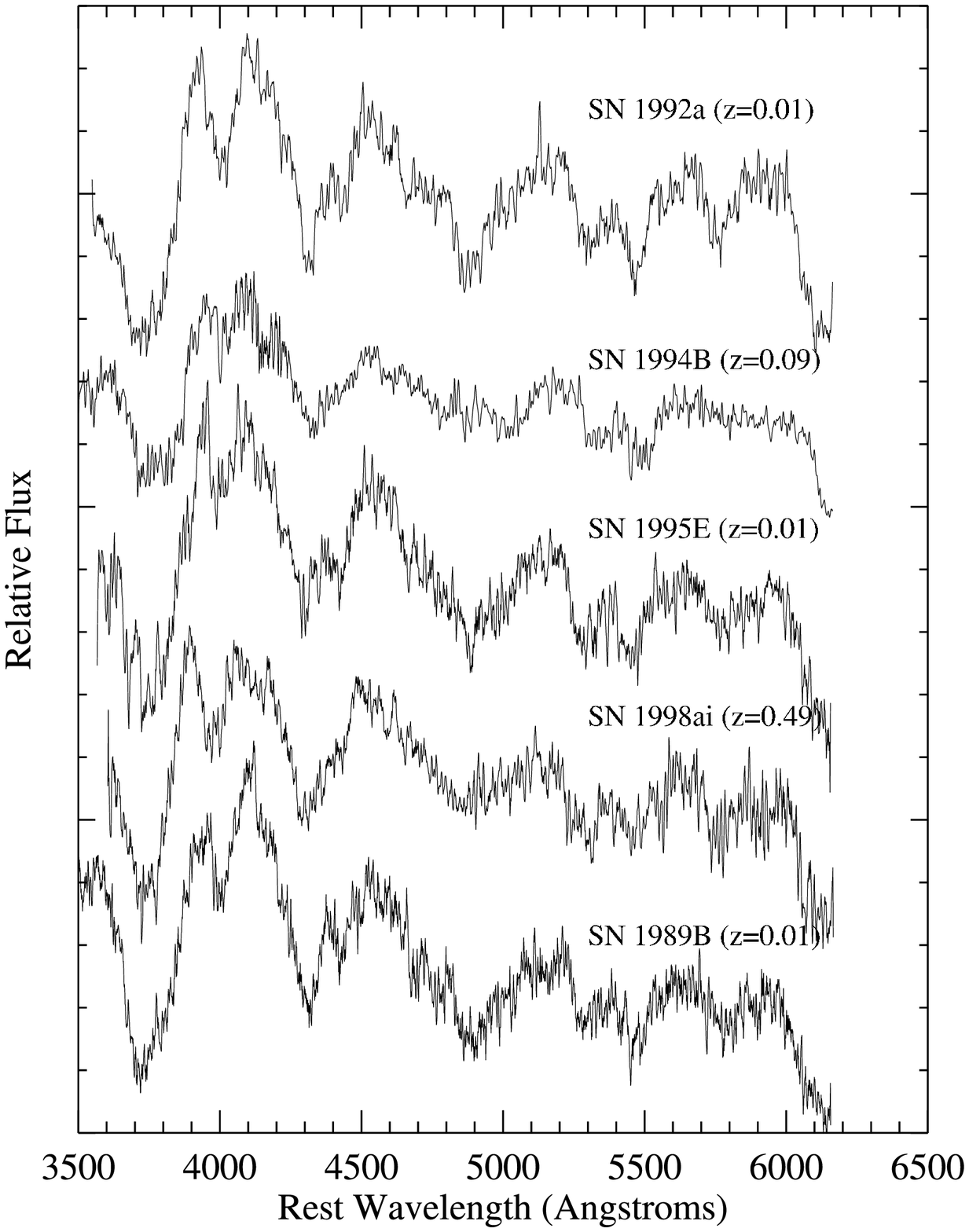,height=3.0truein,angle=0}
}
\hskip 0.0truein
\vbox{\hsize 2.15 truein
\psfig{figure=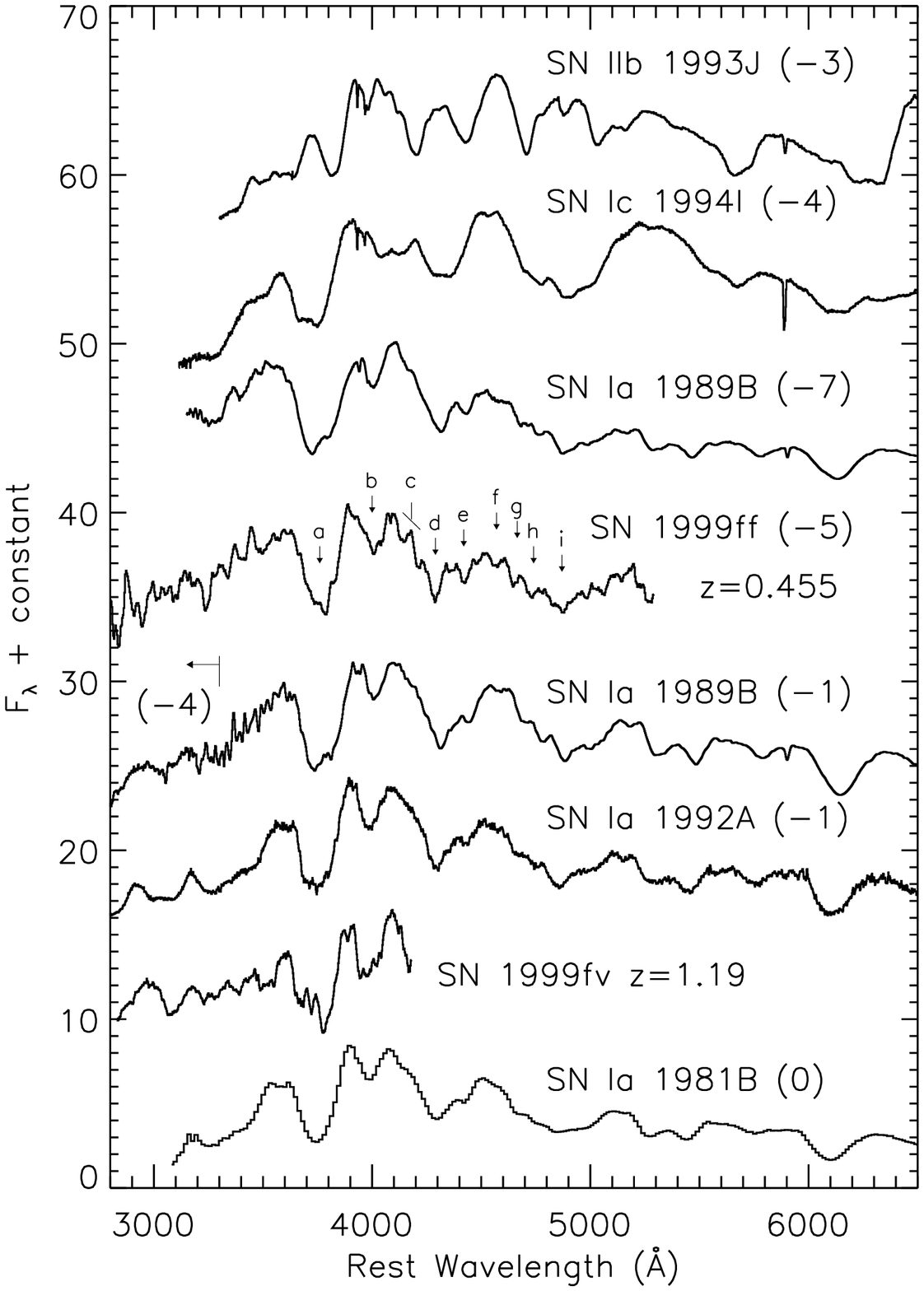,height=3.0truein,angle=0}
}
}

\noindent
{\it Figure 7 (left):} 
Spectral comparison (in $f_{\lambda}$) of SN 1998ai ($z = 0.49$; Keck spectrum)
with low-redshift ($z < 0.1$) SNe~Ia at a similar age ($\sim 5$ days before
maximum brightness), from Riess et al. (1998b).  The spectra of the
low-redshift SNe~Ia were resampled and convolved with Gaussian noise to match
the quality of the spectrum of SN 1998ai. Overall, the agreement in the spectra
is excellent, tentatively suggesting that distant SNe~Ia are physically similar
to nearby SNe~Ia.  SN 1994B ($z = 0.09$) differs the most from the others, and
was included as a ``decoy.''

\smallskip
\noindent
{\it Figure 8 (right):} 
Heavily smoothed spectra of two high-$z$ SNe (SN 1999ff at $z = 0.455$ and SN
1999fv at $z = 1.19$; quite noisy below $\sim$3500~\AA) are presented along
with several low-$z$ SN Ia spectra (SNe 1989B, 1992A, and 1981B), a SN Ib
spectrum (SN 1993J), and a SN~Ic spectrum (SN 1994I); see Filippenko (1997b) for
a discussion of spectra of various types of SNe. The date of the spectra
relative to $B$-band maximum is shown in parentheses after each object's
name. Specific features seen in SN 1999ff and labeled with a letter are
discussed by Coil et al. (2000). This comparison shows that the two high-$z$
SNe are most likely SNe~Ia.

\bigskip

   Although there are no clear signs that cosmic evolution of SNe~Ia seriously
compromises our results, it is wise to remain vigilant for possible problems.
At low redshifts, for example, we already know that {\it some} SNe~Ia don't
conform with the correlation between light curve shape and luminosity.  SN
2000cx in the S0 galaxy NGC 524, for example, has light curves that cannot be
fit well by any of the fitting techniques currently available (Li et al. 2001a;
Filippenko 2003); see Figure 9. Its late-time color is remarkably blue,
inconsistent with the homogeneity described by Phillips et al. (1999). The
spectral evolution of SN 2000cx is peculiar as well (Li et al. 2001a; Branch et
al. 2004a): the photosphere appears to have remained hot for a long time, and
both iron-peak and intermediate-mass elements move at very high velocities.

\medskip

\hbox{
\hskip 1.73truein
\vbox{\hsize 1.0 truein
\psfig{figure=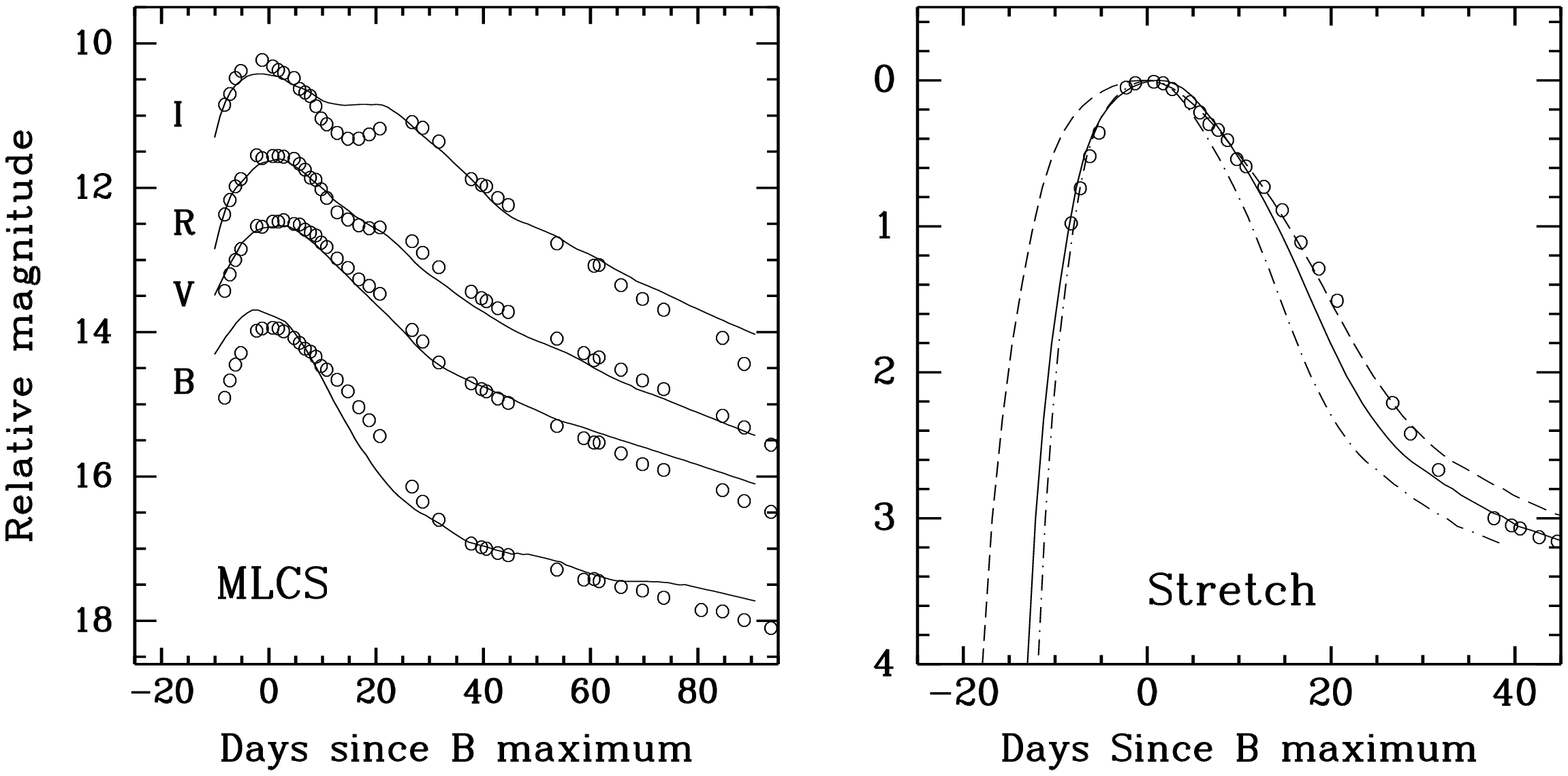,height=0.8truein,angle=-90}
}
}

\bigskip

\noindent 
{\it Figure 9:}
The MLCS fit (Riess et al. 1998b; {\it left panel}) and the stretch method fit
(Perlmutter et al. 1999; {\it right panel}) for SN 2000cx. The MLCS fit is the
worst we had ever seen through year 2000. For the stretch method fit, the solid line
is the fit to all the data points from $t$ = $-$8 to 32 days, the dash-dotted
line uses only the premaximum datapoints, and the dashed line only the
postmaximum datapoints. The three fits give very different stretch factors.
From Li et al. (2001a).

\bigskip

   An even {\it more} peculiar object is SN 2002cx (Li et al. 2003; Filippenko
2003; Branch et al. 2004b). It is spectroscopically bizarre, with extremely low
expansion velocities and almost no evidence for intermediate-mass elements. The
nebular phase was reached incredibly soon after maximum brightness, despite the
low velocity of the ejecta, suggesting that the ejected mass is small. SN
2002cx was subluminous by $\sim 2$ mag at all optical wavelengths relative to
normal SNe~Ia, despite the early-time spectroscopic resemblance to the somewhat
overluminous SN 1991T.  The $R$-band and $I$-band light curves of SN 2002cx are
completely unlike those of normal SNe~Ia. No existing theoretical model
successfully explains all observed aspects of SN 2002cx, though 3D deflagration
models may be best. If there are more strange beasts like SNe 2000cx and 2002cx
at high redshifts than at low redshifts, systematic errors may creep into the
analysis of high-$z$ SNe~Ia.

\subsection{Extinction}
 
   Our SN~Ia distances have the important advantage of including corrections
for interstellar extinction occurring in the host galaxy and the Milky
Way. Extinction corrections based on the relation between SN~Ia colors and
luminosity improve distance precision for a sample of nearby SNe~Ia that
includes objects with substantial extinction (Riess et al. 1996a; Phillips et
al. 1999); the scatter in the Hubble diagram is much reduced.  Moreover, the
consistency of the measured Hubble flow from SNe~Ia with late-type and
early-type hosts (see \S 5.1) shows that the extinction corrections applied to
dusty SNe~Ia at low redshift do not alter the expansion rate from its value
measured from SNe~Ia in low-dust environments.

   In practice, the high-redshift SNe~Ia generally appear to suffer very little
extinction; their $B-V$ colors at maximum brightness are normal, suggesting
little color excess due to reddening. The most detailed available study is that
of the SCP (Sullivan et al. 2003): they found that the scatter in the Hubble
diagram is minimal for SNe~Ia in early-type host galaxies, but increases for
SNe~Ia in late-type galaxies. Moreover, on average the SNe in late-type
galaxies are slightly fainter (by $0.14 \pm 0.09$ mag) than those in early-type
galaxies. Finally, at peak brightness the colors of SNe~Ia in late-type
galaxies are marginally redder than those in early-type galaxies. Sullivan et
al. (2003) conclude that extinction by dust in the host galaxies of SNe~Ia is
one of the major sources of scatter in the high-redshift Hubble diagram.  By
restricting their sample to SNe~Ia in early-type host galaxies (presumably with
minimal extinction), they obtain a very tight Hubble diagram that suggests a
nonzero value for $\Omega_\Lambda$ at the $5\sigma$ confidence level, under the
assumption that $\Omega_{\rm total} = 1$. In the absence of this assumption,
SNe~Ia in early-type hosts still imply that $\Omega_\Lambda > 0$ at nearly the
98\% confidence level. The results for $\Omega_\Lambda$ with SNe~Ia in
late-type galaxies are quantitatively similar, but statistically less secure
because of the larger scatter.

   Riess, Press, \& Kirshner (1996b; see also Phillips et al. 1999) found
indications that the Galactic ratios between selective absorption and color
excess are similar for host galaxies in the nearby ($z \leq 0.1$) Hubble flow.
Yet, what if these ratios changed with lookback time (e.g., Aguirre 1999a)?
Could an evolution in dust-grain size descending from ancestral interstellar
``pebbles'' at higher redshifts cause us to underestimate the extinction?
Large grains would not imprint the reddening signature of typical
interstellar extinction upon which our corrections necessarily rely.

\smallskip

\hbox{
\hskip -0.05truein
\vbox{\hsize 2.2 truein
\psfig{figure=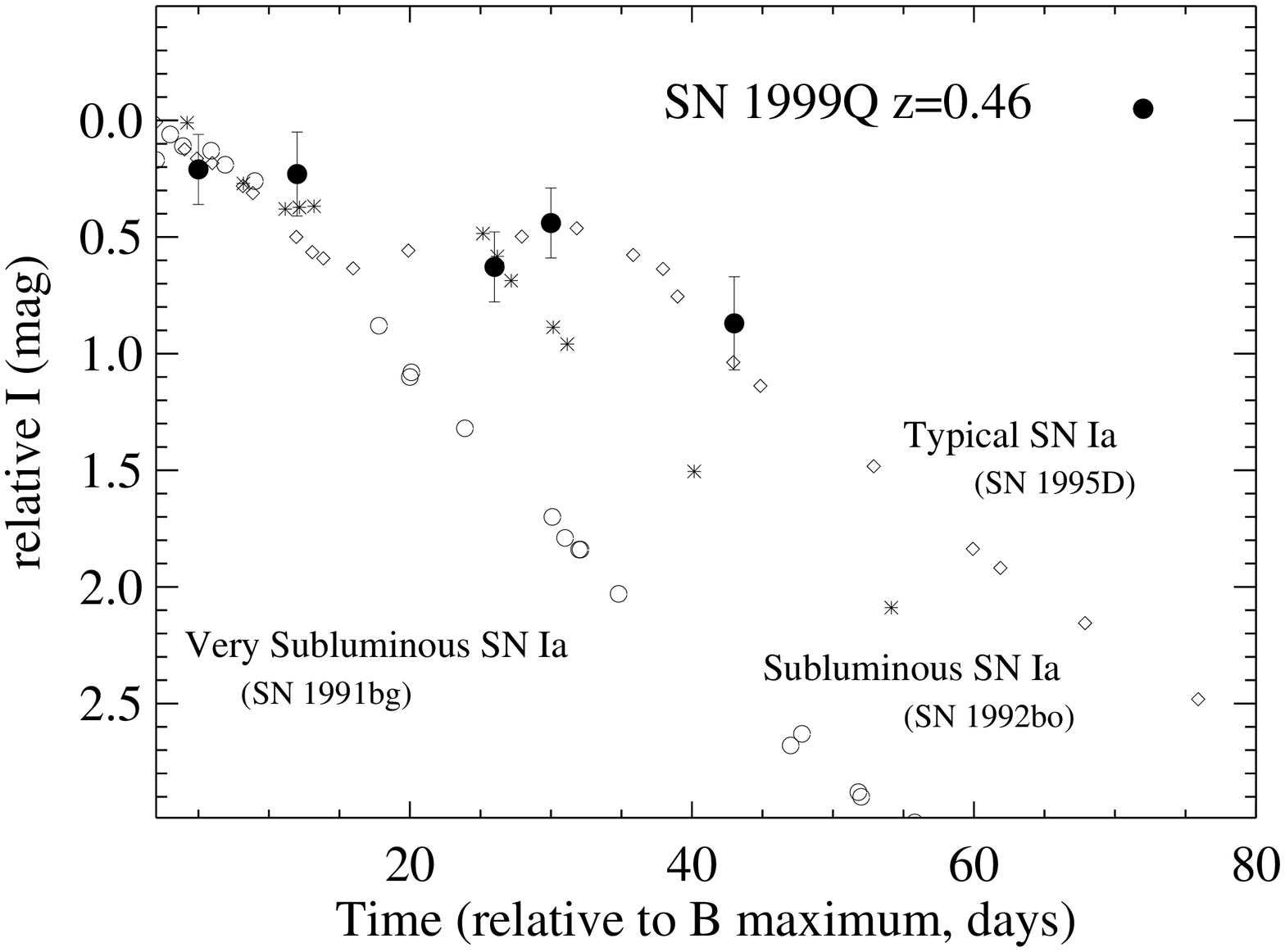,height=2.2truein,angle=90}
}
\hskip -0.1truein
\vbox{\hsize 2.4 truein
\psfig{figure=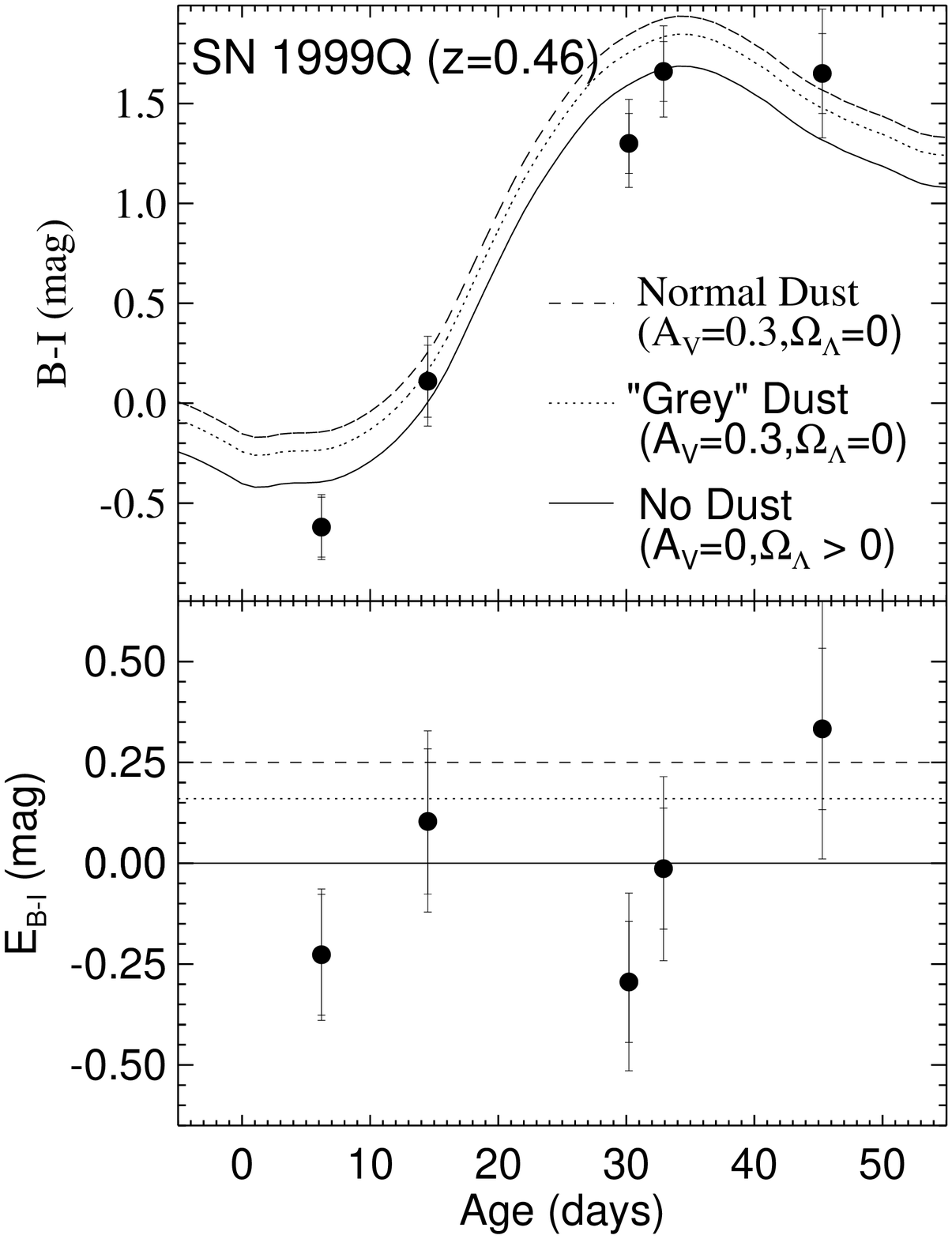,height=2.5truein,angle=0}
}
}

\noindent
{\it Figure 10 (left):} Rest-frame $I$-band (observed $J$-band) light curve of
SN 1999Q ($z = 0.46$, 5 solid points; Riess et al. 2000), and the $I$-band
light curves of several nearby SNe~Ia. Subluminous SNe~Ia exhibit a less
prominent second maximum than do normal SNe~Ia.

\smallskip
\noindent
{\it Figure 11 (right):} Color ($B-I$) and color excess ($E_{B-I}$) for 
SN 1999Q and different dust models (Riess et al. 2000). The data are most 
consistent with no dust and $\Omega_\Lambda > 0$.

\bigskip

   However, viewing our SNe through such gray interstellar grains would also
induce a {\it dispersion} in the derived distances. Using the results of
Hatano, Branch, \& Deaton (1998), Riess et al. (1998b) estimate that the
expected dispersion would be 0.40 mag if the mean gray extinction were 0.25 mag
(the value required to explain the measured MLCS distances without a
cosmological constant).  This is significantly larger than the 0.21 mag
dispersion observed in the high-redshift MLCS distances.  Furthermore, most of
the observed scatter is already consistent with the estimated {\it statistical}
errors, leaving little to be caused by gray extinction.  Nevertheless, if we
assumed that {\it all} of the observed scatter were due to gray extinction, the
mean shift in the SN~Ia distances would be only 0.05 mag.  With the existing
observations, it is difficult to rule out this modest amount of gray
interstellar extinction.

  Gray {\it intergalactic} extinction could dim the SNe without either telltale
reddening or dispersion, if all lines of sight to a given redshift had a
similar column density of absorbing material.  The component of the
intergalactic medium with such uniform coverage corresponds to the gas clouds
producing Lyman-$\alpha$ forest absorption at low redshifts.  These clouds have
individual H~I column densities less than about $10^{15} \, {\rm cm^{-2}}$
(Bahcall et al. 1996).  However, they display low metallicities, typically less
than 10\% of solar. Gray extinction would require larger dust grains which
would need a larger mass in heavy elements than typical interstellar grain size
distributions to achieve a given extinction. It is possible that large dust
grains are blown out of galaxies by radiation pressure, and are therefore not
associated with Lyman-$\alpha$ clouds (Aguirre 1999b).

  But even the dust postulated by Aguirre (1999a,b) and Aguirre \& Haiman
(1999) is not {\it completely} gray, having a size of about 0.1~$\mu$m. We can
test for such nearly gray dust by observing high-redshift SNe~Ia over a wide
wavelength range to measure the color excess it would introduce. If $A_V =
0.25$ mag, then $E(U-I)$ and $E(B-I)$ should be 0.12--0.16 mag (Aguirre
1999a,b). If, on the other hand, the 0.25 mag faintness is due to $\Lambda$,
then no such reddening should be seen.  This effect is measurable using proven
techniques; so far, with just one SN~Ia (SN 1999Q; Figure 11), our results
favor the no-dust hypothesis to better than 2$\sigma$ (Riess et al.  2000).
More work along these lines is in progress.

\subsection{Early Deceleration of the Universe}

   Suppose, however, that for some reason the dust is {\it very} gray, or our
color measurements are not sufficiently precise to rule out Aguirre's (or
other) dust. Or, perhaps some other astrophysical systematic effect is fooling
us, such as possible evolution of the white dwarf progenitors (e.g.,
H\"{o}flich et al. 1998; Umeda et al. 1999), or gravitational lensing
(Wambsganss et al. 1998). The most decisive test to distinguish between
$\Lambda$ and cumulative systematic effects is to examine the {\it deviation}
of the observed peak magnitude of SNe~Ia from the magnitude expected in the
low-$\Omega_M$, zero-$\Lambda$ model. If $\Lambda$ is positive, the deviation
should actually begin to {\it decrease} at $z \approx 1$; we will be looking so
far back in time that the $\Lambda$ effect becomes small compared with
$\Omega_M$, and the Universe is {\it decelerating} at that epoch.  If, on the
other hand, a systematic bias such as gray dust or evolution of the white dwarf
progenitors is the culprit, we expect that the deviation of the apparent
magnitude will continue growing, unless the systematic bias is set up in such
an unlikely way as to mimic the effects of $\Lambda$ (Drell et al. 2000). A
turnover, or decrease of the deviation of apparent magnitude at high redshift,
can be considered almost like a ``smoking gun'' of $\Lambda$ (or, more
generally, of a dark-energy component whose density does not change much
with time).

   In a wonderful demonstration of good luck and hard work, Riess et al. (2001)
report on {\it HST} observations of a probable SN~Ia at $z \approx 1.7$ (SN
1997ff, the most distant SN ever observed) that suggest the expected turnover
is indeed present, providing a tantalizing glimpse of the epoch of
deceleration. (See also Ben\'\i tez et al. 2002, which corrects the observed
magnitude of SN 1997ff for gravitational lensing by foreground galaxies.) SN
1997ff was discovered by Gilliland \& Phillips (1998) in a repeat {\it HST}
observation of the Hubble Deep Field--North, and it was serendipitously
monitored in the infrared with {\it HST}/NICMOS. The peak apparent SN
brightness is consistent with that expected in the decelerating phase of the
concordance cosmological model, $\Omega_M \approx 0.3$, $\Omega_\Lambda \approx
0.7$ (Figure 12). It is inconsistent with gray dust or simple luminosity
evolution, when combined with the data for SNe~Ia at $z \approx 0.5$. 

\smallskip

\hbox{
\hskip 0.25truein
\vbox{\hsize 3.8 truein
\vskip -0.9 truein
\psfig{figure=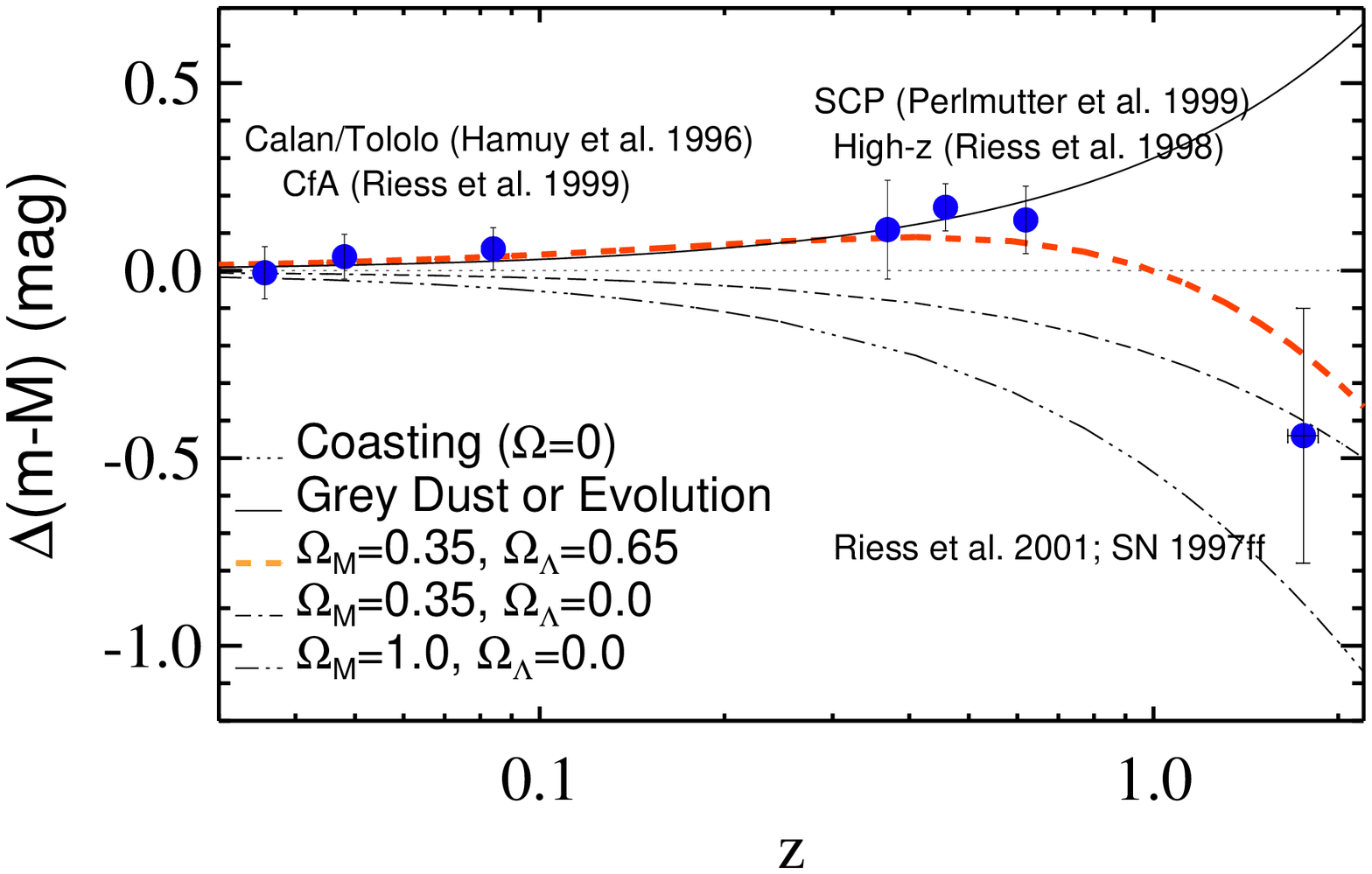,height=4.45truein,angle=0}
\vskip -0.5 truein
}
}

\noindent
{\it Figure 12:} Hubble diagram for SNe~Ia relative to an empty universe
($\Omega = 0$) compared with cosmological and astrophysical models (Riess et
al. 2001). Low-redshift SNe~Ia are from Hamuy et al. (1996a) and Riess et
al. (1999a). The magnitude of SN 1997ff at $z = 1.7$ has been corrected for
gravitational lensing (Ben\'\i tez et al. 2002). The measurements of SN 1997ff
are inconsistent with astrophysical effects that could mimic previous evidence
for an accelerating universe from SNe~Ia at $z \approx 0.5$.
 
\medskip

  On the other hand, it was wise to remain cautious at the time: the error bars
are large, and it is always possible that we are being fooled by this single
object. The HZT and SCP thus started programs to find and measure more SNe~Ia
at such high redshifts ($z > 1$). One promising-looking result was that of
Tonry et al. (2003) and Barris et al. (2004) for the HZT: the deviation of
apparent magnitude from the low-$\Omega_M$, zero-$\Lambda$ model for several
new SNe~Ia at $z \approx 1$ is roughly the same as that at $z \approx 0.5$, in
agreement with expectations based on the results of Riess et al. (2001).

   Inspired by his promising results from SN 1997ff, as well as by the
discovery and study of two distant SNe~Ia (Blakeslee et al. 2003) with the {\it
HST} Advanced Camera for Surveys (ACS), Adam Riess launched an extensive {\it
HST} campaign (the ``Hubble Higher-$z$ Supernova Search'') to closely monitor
$\sim 6$ high-redshift ($z \simgt 1$) SNe~Ia, with the goal of more clearly
detecting the early epoch of deceleration.  A major obstacle was the vast
amount of {\it HST} time required to both discover and obtain follow-up
photometry of the faint SNe, but this was overcome by arranging to
``piggyback'' on the Great Observatories Origins Deep Survey (GOODS), in which
$\sim 400$ {\it HST} orbits were being used to obtain deep, multicolor ACS
images of distant galaxies. The GOODS team obtained their data in several
distinct epochs with separations of 45 days, and with suitable filters,
allowing the discovery of high-redshift SNe (e.g., Giavalisco et al. 2002;
Riess et al. 2004a), generally on the rise.  Results from Jha (2002) suggested
that key light-curve-shape parameters, such as the time of maximum and decline
rate, could be determined from rest-frame $U$-band light curves, corresponding
to observed-frame ACS $z$-band images, allowing for fewer epochs of costly
infrared (NICMOS) observations to measure the SN luminosity and color. This
strategy allowed the requisite follow-up images (e.g., Figure 13) and ACS grism
spectra to be obtained with a relatively modest award of additional {\it HST}
time.

\bigskip
\bigskip
\bigskip
\bigskip
\bigskip
\bigskip

\hbox{
\hskip -0.1truein
\vbox{\hsize 4.5 truein
\vskip -0.9 truein
\psfig{figure=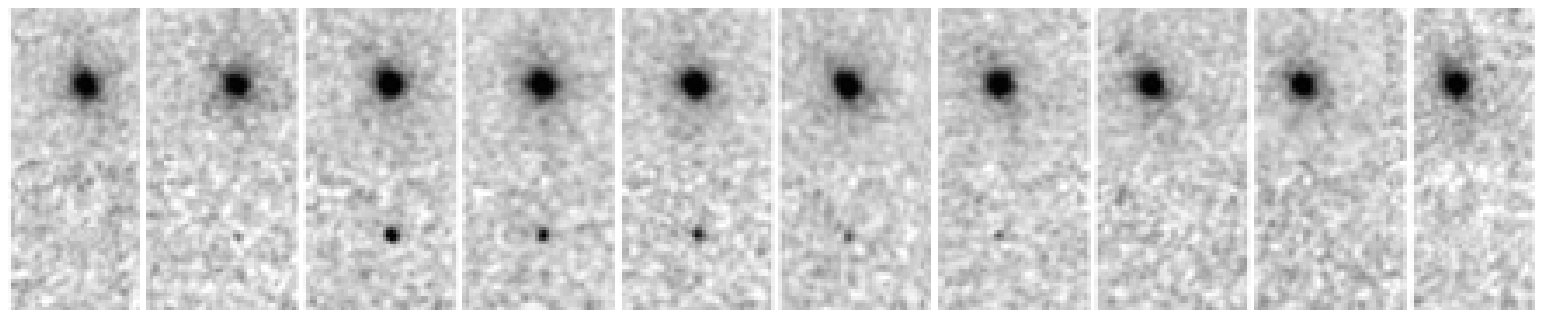,height=0.9truein,angle=0}
\vskip 0.1 truein
}
}

\noindent
{\it Figure 13:} SN 2002hp, a high-redshift supernova from the GOODS program
using {\it HST}. One can see it brightening and subsequently fading with time.
The assumed host galaxy is at the top of each frame.

\medskip

   The program was very successful. Riess et al. (2004b; see also Strolger et
al. 2004) convincingly show that SNe~Ia at $z \simgt 1$ are brighter than
predicted in simple models invoking systematic effects (Fig. 14). Thus, many
SN~Ia evolution and dust models are excluded as alternatives to acceleration
for the observed faintness of $z \approx 0.5$ SNe~Ia. The data demonstrate that
the early universe was indeed apparently decelerating, as expected if $\Lambda$
is a relatively recent effect (Fig. 15).  The expansion of the Universe made a
transition between deceleration and acceleration (Turner \& Riess 2002) at $z
\approx 0.5$, consistent with the new ``concordance cosmology'' of $\Omega_M
\approx 0.27$, $\Omega_\Lambda \approx 0.73$ from the best CMB measurements
(Spergel et al. 2003) and (independently) the complete set of available SN~Ia
data (Riess et al. 2004b).  Finally, Riess et al. (2004b) show that the value
of the dark energy equation-of-state parameter is consistent with $w = -1$, and
that changes in $w$ could not have been very large over the past $\sim 10$
billion years (i.e., $dw/dz = w' \approx 0$, as expected with $\Lambda$). Even
better agreement with the $\Lambda$-model expectations was obtained by Wang \&
Tegmark (2004), who analyzed the data in a slightly different manner. Thus,
although we cannot exclude the possibility that the Universe will recollapse in
the future, such a ``big crunch'' (or ``gnaB giB,'' which is ``Big Bang''
backwards) is unlikely to occur in fewer than 15--30 Gyr (Riess et al. 2004b)
if we adopt the linear potential field of Kallosh \& Linde (2003).

\hbox{
\hskip -0.15truein
\vbox{\hsize 4.0 truein
\psfig{figure=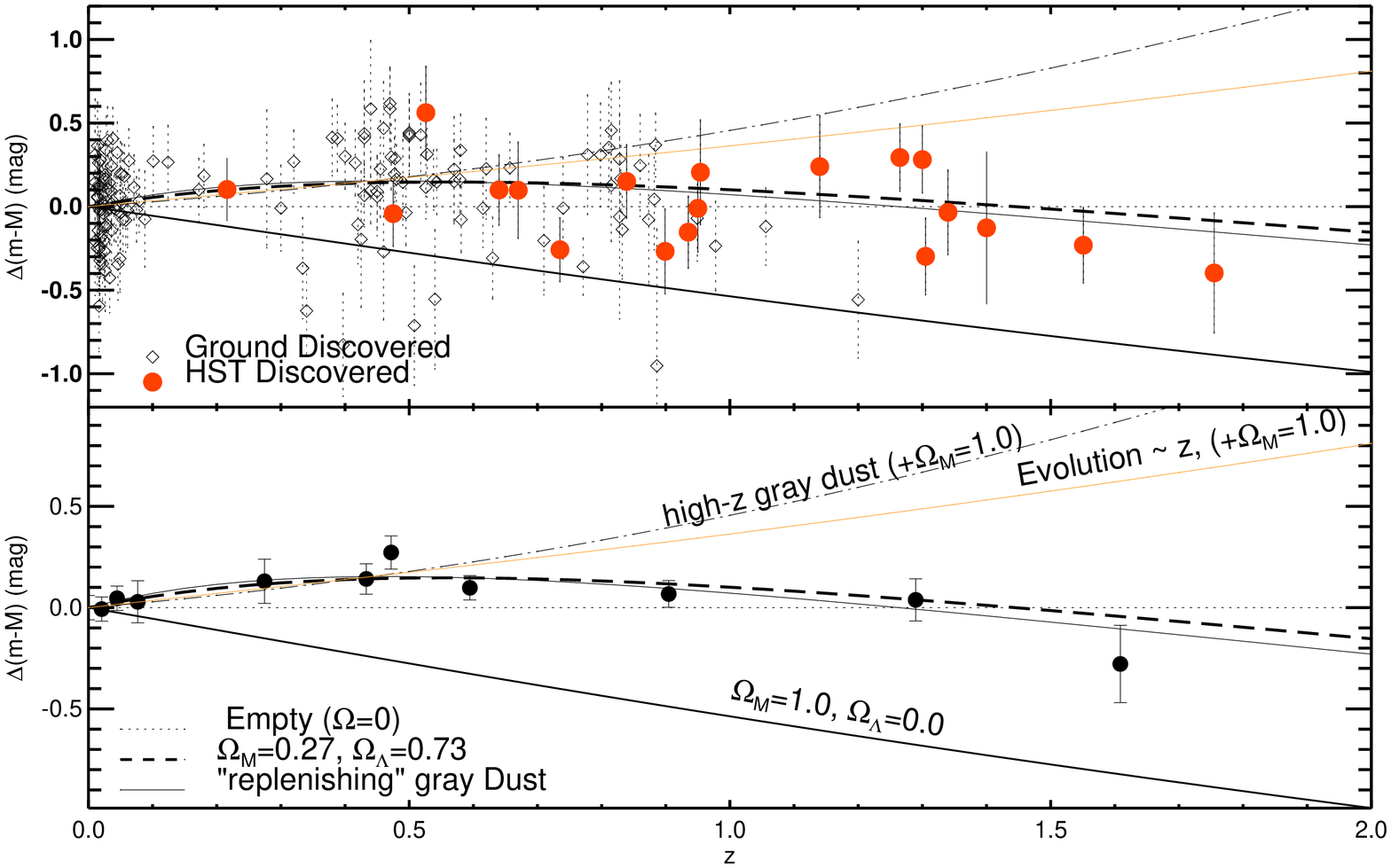,height=4.6truein,angle=90}
}
}

\noindent
{\it Figure 14:} SN~Ia residual Hubble diagram comparing cosmological
models and scenarios for astrophysical dimming (Riess et al. 2004b). {\it Upper
panel:} SNe~Ia from ground-based discoveries are shown as diamonds, {\it
HST}-discovered SNe~Ia are shown as filled symbols. {\it Lower panel:} Weighted
averages in fixed redshift bins, which are given only for illustrative
purposes. Data and models are shown relative to an empty universe model
($\Omega = 0$).

\hbox{
\hskip -.15truein
\vbox{\hsize 4.0 truein
\psfig{figure=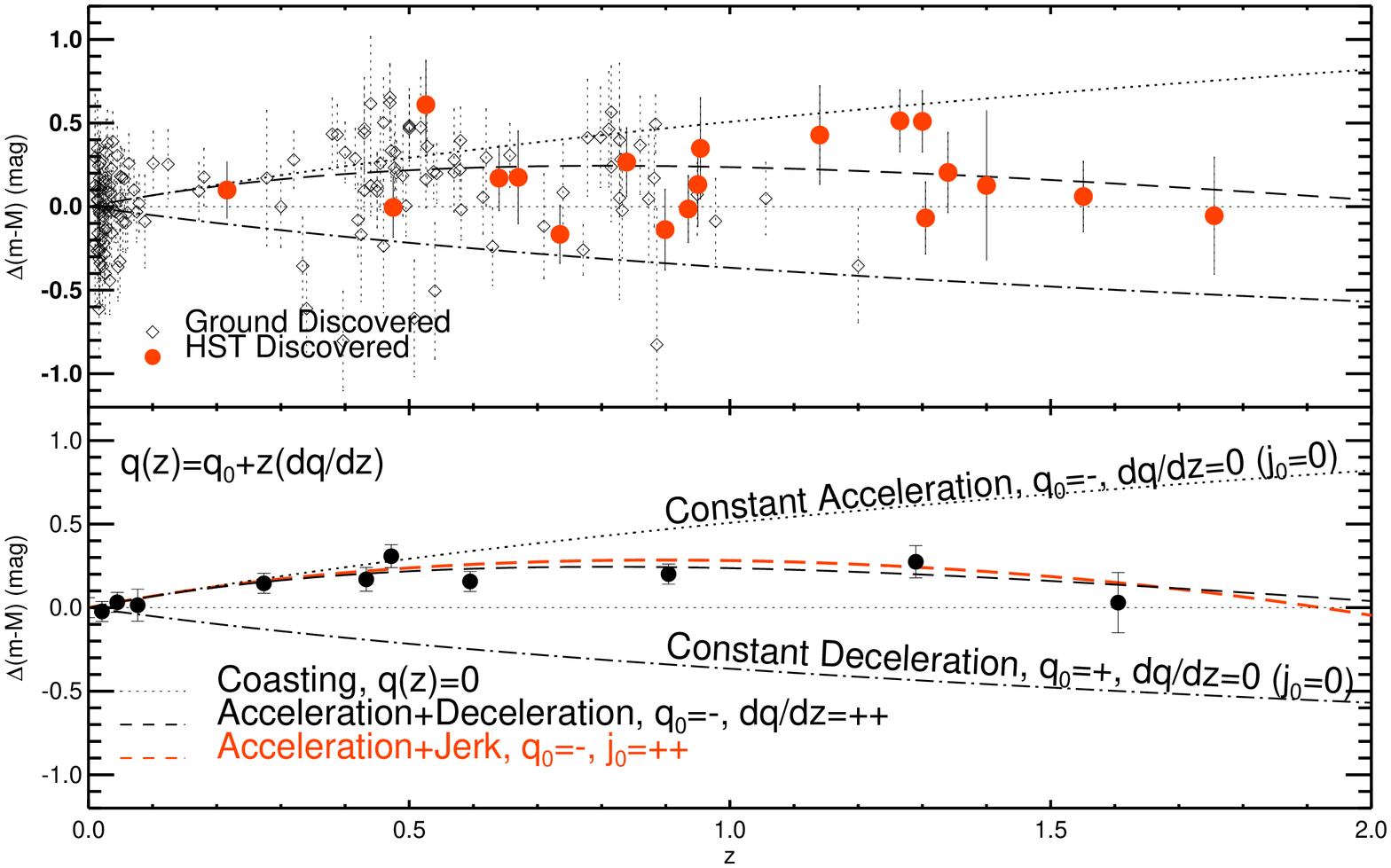,height=4.6truein,angle=90}
}
}

\noindent
{\it Figure 15:} Kinematic SN~Ia residual Hubble diagram (Riess et
al. 2004b). {\it Upper panel:} Symbols as in Figure 14. {\it Lower panel:}
Weighted averages, as in Fig. 14.  Data and kinematic models of the expansion
history are shown relative to an eternally coasting model, $q(z) = 0$. Models
representing specific kinematic scenarios (e.g., ``constant acceleration'') are
illustrated.

\bigskip

   Of course, it is possible to find specific dust or evolution models that are
{\it not} ruled out with our SN~Ia data. For example, a ``replenishing dust''
model represents a constant density of dust that is continually replenished at
precisely the rate at which it is diluted by the expanding universe (Goobar,
Bergstrom, \& Mortsell 2002; Riess et al. 2004b); one could perhaps call this
the ``steady-state dust model.'' Since the dimming is directly proportional to
the distance light traveled and is thus mathematically quite similar to the
effects of a cosmological constant, we cannot discriminate this model from the
$\Lambda$-dominated model in the magnitude-redshift plane. However, the
fine-tuning required of this dust's opacity, replenishing rate, and velocity
($\simgt 1000$ km s$^{-1}$ for it to fill space uniformly without adding
detectable dispersion) makes it unattractive as a simpler alternative to a
cosmological constant. Moreover, the density of intergalactic dust needed to
explain the observed faintness of SNe~Ia at $z \approx 0.5$ exceeds the upper
limit determined by Paerels et al. (2002) from the absence of a detectable
X-ray scattering halo around a single distant quasar.

    Another model with this behavior is evolution that is proportional to
look-back time (Wright 2001). While possible, such dimming behavior, especially
if in the form of luminosity evolution, seems implausible. We may expect
evolution (or dust production) to be coupled to the observed evolution of
stellar populations, galaxies morphologies, sizes, large-scale structure, or
even chemical enrichment. None of these known varieties of evolution are
largely completed by $z = 0.5$ starting from their properties at $z = 0$; quite
the contrary, most of them have hardly begun, looking back to $z = 0.5$.
As mentioned previously, a strong empirical argument against recent luminosity
evolution is the independence of SN Ia distance measurements on gross host
morphology (Riess et al. 1998b; Sullivan et al. 2003). The range of 
progenitor formation environments spanned by SNe Ia in early-type and
late-type hosts greatly exceeds any evolution in the mean host properties
between $z = 0$ and $z = 0.5$. In the end, however, the only ``proof'' against
astrophysical contamination of the cosmological signal from SNe~Ia is to test
the results against those of other observations, independent of SNe~Ia.

\subsection{Type Ia Supernova Rates and Progenitors}

   The rate of SNe~Ia as a function of redshift is important to know for a
number of reasons. For example, it affects the derived chemical evolution of
galaxies (e.g., Maoz \& Gal-Yam 2004; Matteucci \& Recchi 2001, and references
therein), and it can be used to set constraints on the star formation history of
the Universe (Gal-Yam \& Maoz 2004). Moreover, the distribution of ``delay times''
between the formation of the progenitor star and the explosion of the supernova
can be determined from an accurate census of the SN~Ia rate at different cosmic
times, thereby providing clues to the physical nature of the progenitors of
SNe~Ia (Madau, Della Valle, \& Panagia 1998; Dahlen \& Fransson 1999; Yungelson
\& Livio 2000).  There have been many measurements of the rate of very nearby
SNe~Ia (Cappellaro, Evans, \& Turatto 1999, and references therein), and at $z
\approx 0.1$ several studies have been conducted (Hardin et al. 2000; Strolger
2003; Reiss 2000). These form the basis of comparison for the measurements at
higher redshifts.

   From observations of three SNe~Ia at $z \approx 0.4$, Pain et al. (1996)
estimated a rate of SNe~Ia equivalent to 34 yr$^{-1}$ deg$^{-2}$, with a
1$\sigma$ uncertainty of a factor of $\sim 2$, for objects found in the range
of $21.3 < R < 22.3$ mag.  A more recent estimate by Pain et al. (2002a) is
$(1.53 \pm 0.3) \times 10^{-4} \; h^3$ Mpc$^{-3}$ yr$^{-1}$ at a mean redshift
of 0.55 (where $h \equiv H_0/(100~{\rm km}~{\rm s}^{-1}~{\rm
Mpc}^{-1}$). Cappellaro et al. (1999) report a nearby SN~Ia rate of $0.36 \pm
0.11\,h^2$ SNu (where 1 SNu $\equiv 10^{10}\,L_{B_\odot}$ per century), or
$(0.79 \pm 0.24) \times 10^{-4} \; h^3$ Mpc$^{-3}$ yr$^{-1}$, using a quite
uncertain local luminosity density of $\rho = 2.2 \times 10^{8} \; h \;
L_{B_\odot}$ Mpc$^{-3}$.  Pain et al. (2002a) claim to see a very modest
increase in the rate of SNe~Ia with redshift, perhaps tracking the star
formation rate which Wilson et al. (2002) estimate as being proportional to
$(1+z)^{1.7}$ in a flat universe with $\Omega_M = 0.3$.

   Based on a new sample of SNe~Ia observed during the 1999 HZT campaign, an
independent estimate of the SN~Ia rate at a mean redshift of 0.46 was made by
Tonry et al. (2003): $(1.4 \pm 0.5) \times 10^{-4} \; h^3$ Mpc$^{-3}$
yr$^{-1}$. This is in excellent agreement with the results of Pain et
al. (2002a), and it is not inconsistent with the local rates from Cappellaro et
al. (1999), particularly given the uncertainty in the local luminosity
density. On the other hand, from a different sample of SNe~Ia (see below),
Dahlen et al. (2004) find a somewhat higher rate of SNe~Ia at $z \approx 0.4$
and suggest that the previously published rates may have been underestimated
due to systematic effects.  Indeed, B. Barris (2004, private communication) has
found evidence that the true rate of SNe~Ia in the 2001 HZT campaign is higher
than that reported by Barris et al. (2004); similar biases may have affected
the 1999 HZT campaign and other previous searches.

   Although Tonry et al. (2003) were incomplete in their counts of SNe~Ia at $z
\approx 1$, they do not believe that the rate of SNe~Ia closely tracks the
star formation rate.  Application of the star formation rate from Wilson et
al. (2002) would suggest that the SN~Ia rate at $z \approx 1.1$ should be three
times as great as occurs locally and nearly twice as great as the rate at $z =
0.46$. In this case, Tonry et al. (2003) should have discovered 16 SNe~Ia
deg$^{-2}$ in their search, yet they only found 4 SNe~Ia deg$^{-2}$. It is
their impression that the constant rate per volume is closer to the truth.

   The more recent results of Dahlen et al. (2004) conflict with this
conclusion. They determine the rate of SNe~Ia as a function of redshift by
using data from the Hubble Higher-$z$ Supernova Search of Riess et al. (2004b),
with a redshift range of $0.2 < z < 1.6$. The resulting SN~Ia rate at $z
\approx 1$ is a factor of 3--5 higher than previous estimates made at lower
redshifts ($z < 0.5$), presumably because the star formation rate at $z > 1$
was substantially higher than that in the past few Gyr.  Moreover, their data
suggest that at even higher redshifts, $z > 1$, the rate of SNe~Ia begins to
decrease. They find that the delay time (from progenitor star formation to
SN~Ia explosion) is likely to be substantial, $\simgt 2$ Gyr. In addition,
assuming a Salpeter (1955) initial mass function and a SN~Ia progenitor
main-sequence mass range of 3--8~$M_\odot$, it appears that 5--7\% of the white
dwarfs formed from these progenitor stars eventually become SNe~Ia (but fewer
than 1\% of {\it all} white dwarfs eventually become SNe~Ia).

   Strolger et al. (2004) use the above results, together with additional
modeling, to set more quantitative constraints on the nature of the progenitors
of SNe~Ia.  Specifically, they use a Bayesian maximum likelihood test to
determine the most likely range of delay times that best reproduces the
observed redshift distribution of SNe~Ia. They find that models requiring a
large fraction of ``prompt'' (less than 2 Gyr) SNe~Ia poorly reproduce the
observed redshift distribution and are rejected at $> 99$\% confidence. Thus,
SNe~Ia cannot generally be prompt events, nor can they be expected to closely
follow the star formation rate history. Instead, Gaussian models best fit the
observed data for mean delay times in the range of 3--4 Gyr. This may be most
consistent with single-degenerate systems in which the white dwarf accretes
from a main-sequence companion or from a somewhat evolved companion (Livio
2001), although certain types of double-degenerate models are not yet
eliminated.  Tests conducted by Gal-Yam \& Maoz (2004) also conclude that the
characteristic delay times of SNe~Ia should be large ($>$1--2 Gyr) for similar
assumed models of the star formation rate history, but the results are not as
definitive as those of Strolger et al. (2004) because they are based on SN~Ia
rates derived from more limited SCP data (Pain et al. 2002a).

\subsection{Measuring the Dark Energy Equation-of-State Parameter}
 
   Every energy component in the Universe can be parameterized by the way its
density varies as the Universe expands (scale factor $a$), with $\rho \propto
a^{-3(1+w)}$, and $w$ is the component's equation-of-state parameter, $w =
P/(\rho c^2)$, where $P$ is the pressure exerted by the component.  So for
matter, $w=0$, while an energy component that does not vary with scale factor
has $w=-1$, as in the cosmological constant $\Lambda$. Quintessence models have
$w \neq -1$, and generally $dw/dz \neq 0$). Some really strange energies may
have $w < -1$: their density increases with time (Carroll, Hoffman, \& Trodden
2003), leading to a ``Big Rip'' in which progressively smaller bound systems
get torn apart! [Riess et al. (2004b) and Caldwell et al. (2003) estimate that
such a fate will not occur sooner than $\sim 20$ Gyr from now, if ever.]
Clearly, a good estimate of $w$ becomes the key to differentiating between
models.

The CMB observations imply that the geometry of the universe is close to flat,
so the energy density of the dark component is simply related to the matter
density by $\Omega_x = 1 - \Omega_M$.  This allows the luminosity distance as a
function of redshift to be written as
$$D_L(z)\;
=\; {c(1+z)\over{H_0}}\int_0^{z}{[1+\Omega_x((1+{\rm z})^{3w}-1)]
^{-{1/2}}\over(1+{\rm z})^{{3/2}}}\;  {\rm dz} \; ,$$
showing that the dark energy density and equation of state directly influence
the apparent brightness of standard candles. As demonstrated graphically in
Figure 16, SNe~Ia observed over a wide range of redshifts can 
constrain the dark energy parameters to a cosmologically interesting accuracy.

But there are two major problems with using SNe~Ia to measure $w$.  First,
systematic uncertainties in SN~Ia peak luminosity limit how well $D_L(z)$ can
be measured. While statistical uncertainty can be arbitrarily reduced by
finding thousands of SNe~Ia, intrinsic SN properties such as evolution and
progenitor metallicity, and observational limits like photometric calibrations
and K-corrections, create a systematic floor that cannot be decreased by sheer
force of numbers. We expect that systematics can be controlled to at best 3\%,
with considerable effort.

Second, SNe at $z > 1.0$ are very hard to discover and study from the
ground. As discussed above, both the HZT and the SCP have found a few SNe~Ia at
$z > 1.0$, but the numbers and quality of these light curves are insufficient
for a $w$ measurement. Large numbers of SNe~Ia at $z > 1.0$ are best left to a
wide-field optical/infrared imager in space, such as the proposed {\it
Supernova/ Acceleration Probe} ({\it SNAP}; Nugent et al. 2001) satellite.

Fortunately, an interesting measurement of $w$ can be made at present.  The
current values of $\Omega_M$ from many methods (most recently {\it WMAP}: 0.27;
Spergel et al. 2003) make an excellent substitute for those expensive SNe at $z
> 1.0$.  Figure 16 shows that a SN~Ia sample with a maximum redshift of $z =
0.8$, combined with the current 10\%\ error on $\Omega_M$, will do as well as a
SN~Ia sample at much higher redshifts. Within a few years, the Sloan Digital
Sky Survey and {\it WMAP}\ will solidify the estimate of $\Omega_M$ and sharpen
$w$ further.

   Both the SCP and the HZT are involved in multi-year programs to discover and
monitor hundreds of SNe~Ia for the purpose of measuring $w$. For example, the
HZT's project, ESSENCE (Equation of State: SupErNovae trace Cosmic Expansion),
is designed to discover 200 SNe~Ia evenly distributed in the $0.2 < z < 0.8$
range (Smith et al. 2002; Garnavich et al. 2002;
http://www.ctio.noao.edu/wproject).  The CTIO 4-m telescope and mosaic camera
are being used to find and follow the SNe by imaging on every other dark night
for several consecutive months of the year. Keck and other large telescopes are
being used to get the SN spectra and redshifts.  Project ESSENCE will
eventually provide an estimate of $w$ to an accuracy of $\sim$10\% (Figure
17). Even larger numbers of high-redshift SNe~Ia are being found during the
ongoing CFHT Legacy Survey (http://www.cfht.hawaii.edu/Science/CFHLS; Pain et
al. 2002b), providing an independent sample with which to measure the value of
$w$.  Within the next few years, Pan-STARRS (Kaiser et al. 2002;
http://poi.ifa.hawaii.edu) should discover and follow thousands of SNe~Ia.

\bigskip
\smallskip

\hbox{
\hskip -0.3truein
\vbox{\hsize 1.7 truein
\psfig{figure=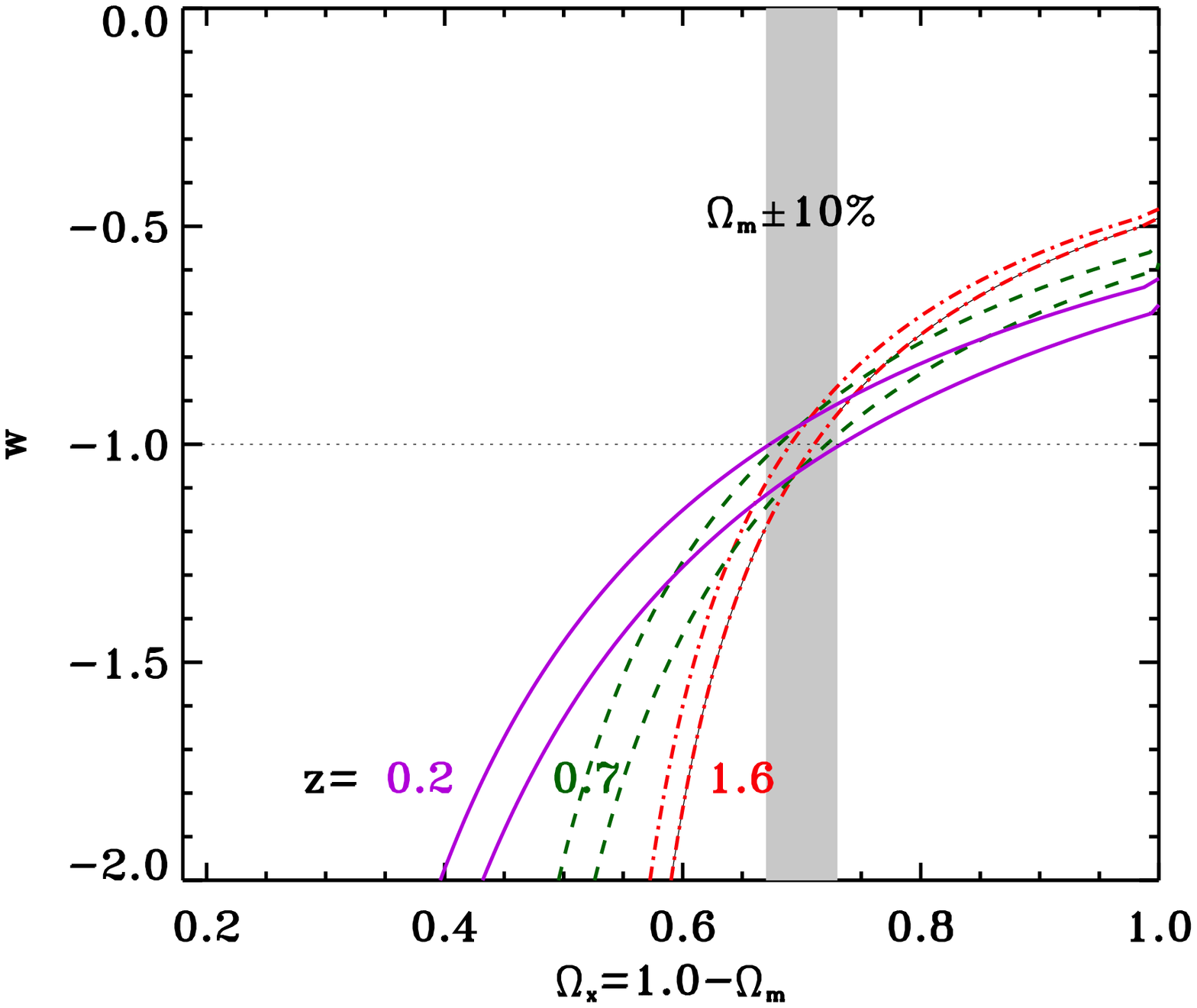,height=1.7truein,angle=0}
}
\hskip +0.7truein
\vbox{\hsize 1.7 truein
\psfig{figure=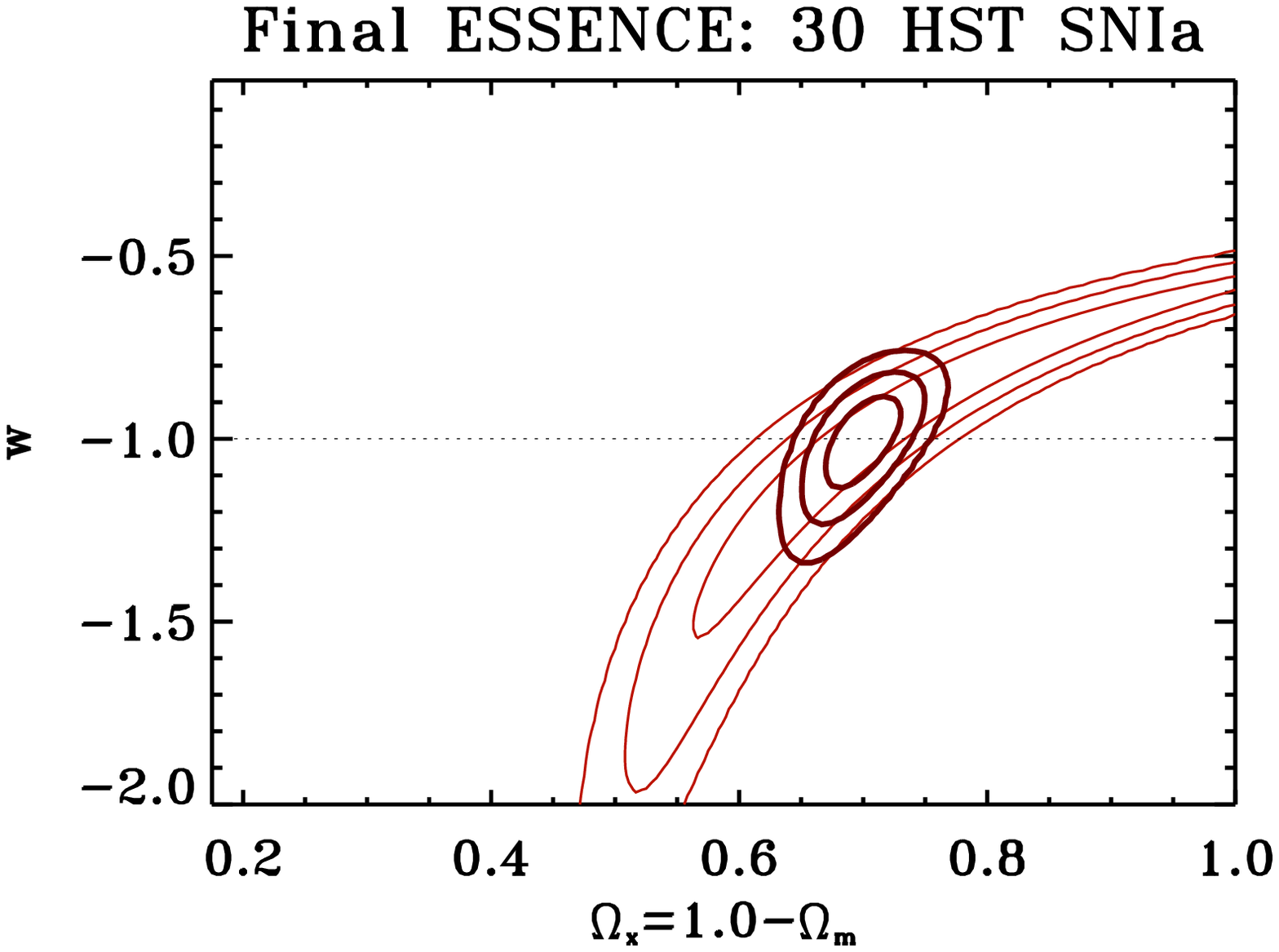,height=1.7truein,angle=0}
}
}

\bigskip
\medskip

\noindent
{\it Figure 16 (left):} Constraints on $\Omega_x$ and $w$ from SN data sets
collected at $z=0.2$ (solid lines), $z=0.7$ (dashed lines), and $z=1.6$
(dash-dot lines). The shaded area indicates how an independent estimate of
$\Omega_M$ with a 10\%\ error can help constrain $w$.

\medskip
\noindent
{\it Figure 17 (right):} Expected constraints on $w$ with the desired final
ESSENCE data set of 200 SNe~Ia, 30 of which (in the redshift range $0.6 < z <
0.8$) are to be observed with {\it HST}. The thin lines are for SNe alone while
the thick lines assume an uncertainty in $\Omega_M$ of 7\%. The final ESSENCE
data will constrain the value of $w$ to $\sim$10\%.

\bigskip

   Farther in the future, the plethora of SNe~Ia to be found and studied by the
proposed {\it SNAP} satellite (Nugent et al. 2001), the NASA/DOE Joint Dark
Energy Mission (JDEM), the Large-area Synoptic Survey Telescope (the ``Dark
Matter Telescope''; Tyson \& Angel 2001), and similar large-scale projects
could reveal whether the value of $w$ depends on redshift, and hence should
give additional constraints on the nature of the dark energy. High-redshift
surveys of galaxies such as DEEP2 (Davis et al.  2001), as well as space-based
missions to map the CMB ({\it Planck}), should provide additional evidence for
(or against) $\Lambda$. Observational cosmology promises to remain exciting for
quite some time!


\begin{acknowledgments}

    I thank all of my HZT collaborators for their contributions to our team's
research, and members of the SCP for their seminal complementary work on the
accelerating Universe. My group's work at U.C. Berkeley has been supported by
NSF grants AST--9987438, AST--0206329, and AST-0307894, as well as by grants
GO--7505, GO/DD--7588, GO--8177, GO--8641, GO--9118, and GO--9352 from the
Space Telescope Science Institute, which is operated by the Association of
Universities for Research in Astronomy, Inc., under NASA contract
NAS~5--26555. Many spectra of high-redshift SNe were obtained at the W. M. Keck
Observatory, which is operated as a scientific partnership among the California
Institute of Technology, the University of California, and NASA; the
observatory was made possible by the generous financial support of the
W. M. Keck Foundation. KAIT has received donations from Sun Microsystems, Inc.,
the Hewlett-Packard Company, AutoScope Corporation, Lick Observatory, the
National Science Foundation, the University of California, and the Sylvia and
Jim Katzman Foundation. I am grateful to the editors of this volume for their
incredible patience while waiting for my review.

\end{acknowledgments}


\begin{chapthebibliography}{}


\bibitem{afs04}
Afshordi, N., Loh, Y.-S., \& Strauss, M. S. 2004, {\it Phys. Rev. D}, 
   {\bf 69}, 083524

\bibitem{agu99a}
Aguirre, A. N. 1999a, {\it ApJ}, {\bf 512}, L19

\bibitem{agu99b}
Aguirre, A. N. 1999b, {\it ApJ}, {\bf 525}, 583

\bibitem{agu99c}
Aguirre, A. N., \& Haiman, Z. 1999, {\it ApJ}, {\bf 525}, 583

\bibitem{ald00}
Aldering, G., Knop, R., \& Nugent, P. 2000, {\it AJ}, {\bf 119}, 2110

\bibitem{bah96}
Bahcall, J. N., et al. 1996, {\it ApJ}, {\bf 457}, 19

\bibitem{bah99}
Bahcall, N. A., Ostriker, J. P., Perlmutter, S., \& Steinhardt, P. J. 1999, 
   {\it Science}, {\bf 284}, 1481

\bibitem{bal00}
Balbi, A., et al. 2000, {\it ApJ}, {\bf 545}, L1

\bibitem{bar00}
Barber, A. J., Thomas, P. A., Couchman, H. M. P., \& Fluke, C. J. 2000, 
  {\it MNRAS}, {\bf 319}, 267

\bibitem{bar04}
Barris, B., et al. 2004, {\it ApJ}, {\bf 602}, 571

\bibitem{ben02}
Ben\'\i tez, N., Riess, A., Nugent, P., Dickinson, M.,
   Chornock, R., \& Filippenko, A. V. 2002, {\it ApJ}, {\bf 577}, L1

\bibitem{bla03}
Blakeslee, J. P., et al. 2003, {\it ApJ}, {\bf 589}, 693

\bibitem{bou04}
Boughn, S., \& Crittenden, R. 2004, {\it Nature}, {\bf 427}, 45

\bibitem{bra81}
Branch, D. 1981, {\it ApJ}, {\bf 248}, 1076

\bibitem{bra98}
Branch, D. 1998, {\it ARA\&A}, {\bf 36}, 17

\bibitem{bra04}
Branch, D., Baron, E., Thomas, R. C., Kasen, D., Li, W., \&
   Filippenko, A. V. 2004b, {\it PASP}, in press (astro-ph/0408130)

\bibitem{bfn93}
Branch, D., Fisher, A., \& Nugent, P. 1993, {\it AJ}, {\bf 106}, 2383

\bibitem{bra93}
Branch, D., \& Miller, D. L. 1993, {\it ApJ}, {\bf 405}, L5 

\bibitem{bra96}
Branch, D., Romanishin, W., \& Baron, E. 1996, {\it ApJ}, {\bf 465}, 73 
    (erratum: {\bf 467}, 473)

\bibitem{bra92}
Branch, D., \& Tammann, G. A. 1992, {\it ARA\&A}, {\bf 30}, 359

\bibitem{bra04}
Branch, D., et al. 2004a, {\it ApJ}, {\bf 606}, 413

\bibitem{cal98}
Caldwell, R. R., Dav\'e, R., \& Steinhardt, P. J. 1998, {\it Ap\&SS}, 
   {\bf 261}, 30

\bibitem{cal03}
Caldwell, R. R., Kamionkowski, M., \& Weinberg, N. N. 2003, {\it Phys. Rev.
  Lett.}, {\bf 91}, 71301

\bibitem{cap99}
Cappellaro, E., Evans, R., \& Turatto, M. 1999, {\it A\&A}, {\bf 351}, 459

\bibitem{cap97}
Cappellaro, E., Turatto, M., Tsvetkov, D. Yu., Bartunov, O. S.,
  Pollas, C., Evans, R., \& Hamuy, M. 1997, {\it A\&A}, {\bf 322}, 431

\bibitem{car03}
Carroll, S. M., Hoffman, M., \& Trodden, M. 2003, {\it Phys. Rev. D}, 
   {\bf 68}, 023509

\bibitem{car92}
Carroll, S. M., Press, W. H., \& Turner, E. L. 1992, {\it ARA\&A}, 
  {\bf 30}, 499

\bibitem{cha98}
Chaboyer, B., Demarque, P., Kernan, P. J., \& Krauss, L. M.  1998, 
   {\it ApJ}, {\bf 494}, 96

\bibitem{coi00}
Coil, A. L., et al. 2000, {\it ApJ}, {\bf 544}, L111

\bibitem{cow97}
Cowan, J. J., McWilliam, A., Sneden, C., \& Burris, D. L. 1997, 
    {\it ApJ}, {\bf 480}, 246

\bibitem{dah99}
Dahlen, T., \& Fransson, C. 1999, {\it A\&A}, {\bf 350}, 349

\bibitem{dah04}
Dahlen, T., et al. 2004, {\it ApJ}, {\bf 613}, 189

\bibitem{dav01}
Davis, M., Newman, J. A., Faber, S. M., \& Phillips, A. C.  2001, in 
   {\it Deep Fields}, ed. S. Cristiani, A. Renzini, \& R. E. Williams 
   (Berlin: Springer), 241

\bibitem{deb00}
de Bernardis, P., et al. 2000, {\it Nature}, {\bf 404}, 955

\bibitem{deb02}
de Bernardis, P., et al. 2002, {\it ApJ}, {\bf 564}, 559

\bibitem{dre00}
Drell, P. S., Loredo, T. J., \& Wasserman, I. 2000, {\it ApJ}, 
   {\bf 530}, 593

\bibitem{efs99}
Efstathiou, G., et al. 1999, {\it MNRAS}, {\bf 303}, L47

\bibitem{efs02}
Efstathiou, G., et al. 2002, {\it MNRAS}, {\bf 330}, L29

\bibitem{eis98}
Eisenstein, D. J., Hu, W., \& Tegmark, M. 1998, {\it ApJ}, 
   {\bf 504}, L57

\bibitem{fal99}
Falco, E. E., et al. 1999, {\it ApJ}, {\bf 523}, 617

\bibitem{avf97a}
Filippenko, A. V. 1997a, in {\it Thermonuclear Supernovae}, 
 ed. P. Ruiz-Lapuente et al. (Dordrecht: Kluwer), 1

\bibitem{avf97b}
Filippenko, A. V. 1997b, {\it ARA\&A}, {\bf 35}, 309

\bibitem{avf01}
Filippenko, A. V. 2001, {\it PASP}, {\bf 113}, 1441

\bibitem{avf03}
Filippenko, A. V. 2003, in {\it From Twilight to Highlight: The 
   Physics of Supernovae}, ed.
   W. Hillebrandt \& B. Leibundgut (Berlin: Springer-Verlag), 171

\bibitem{fil01}
Filippenko, A. V., Li, W. D., Treffers, R. R., \& Modjaz, M. 2001, in 
  {\it Small-Telescope Astronomy on Global Scales}, ed. W. P. Chen, 
  C. Lemme, \& B. Paczy\'{n}ski (San Francisco: ASP), 121

\bibitem{avf98}
Filippenko, A. V., \& Riess, A. G. 1998, {\it Phys. Rep.}, {\bf 307}, 31

\bibitem{avf92a}
Filippenko, A. V., et al. 1992a, {\it AJ}, {\bf 104}, 1543

\bibitem{avf92b}
Filippenko, A. V. 1992b, {\it ApJ}, {\bf 384}, L15

\bibitem{fol03}
Foley, R. J., et al. 2003, {\it PASP}, {\bf 115}, 1220

\bibitem{for93}
Ford, C. H., et al. 1993, {\it AJ}, {\bf 106}, 1101

\bibitem{fos04}
Fosalba, P., et al. 2003, {\it ApJ}, {\bf 597}, L89

\bibitem{fre01}
Freedman, W., et al. 2001, {\it ApJ}, {\bf 553}, 47

\bibitem{gal04}
Gal-Yam, A., \& Maoz, D. 2004, {\it MNRAS}, {\bf 347}, 942

\bibitem{gar98a}
Garnavich, P., et al. 1998a, {\it ApJ}, {\bf 493}, L53

\bibitem{gar98b}
Garnavich, P., et al. 1998b, {\it ApJ}, {\bf 509}, 74 

\bibitem{gar02}
Garnavich, P., et al. 2002, {\it BAAS}, {\bf 34}, 1233

\bibitem{Gia02}
Giavalisco, M., et al. 2002, {\it IAUC 7981}

\bibitem{Gib00}
Gibson, B. K., et al. 2000, {\it ApJ}, {\bf 529}, 723

\bibitem{gil98}
Gilliland, R. L., \& Phillips, M. M. 1998, {\it IAUC 6810}

\bibitem{gol98}
Goldhaber, G., \& Perlmutter, S. 1998, {\it Phys. Rep.}, {\bf 307}, 325

\bibitem{gol97}
Goldhaber, G., et al. 1997, in {\it Thermonuclear Supernovae}, 
   ed. P. Ruiz-Lapuente et al. (Dordrecht: Kluwer), 777

\bibitem{gol98a}
Goldhaber, G., et al. 1998a, {\it BAAS}, {\bf 30}, 1325

\bibitem{gol98b}
Goldhaber, G., et al. 1998b, in {\it Gravity: From the Hubble 
   Length to the Planck Length}, SLAC Summer Institute 
   (Stanford, CA: SLAC)

\bibitem{gol01}
Goldhaber, G., et al. 2001, {\it ApJ}, {\bf 558}, 359

\bibitem{goo02}
Goobar, A., Bergstrom, L., \& Mortsell, E. 2002, {\it A\&A}, {\bf 384}, 1

\bibitem{goo95}
Goobar, A., \& Perlmutter, S. 1995, {\it ApJ}, {\bf 450}, 14

\bibitem{gra97}
Gratton, R. G., Fusi Pecci, F., Carretta, E., Clementini, G.,
  Corsi, C. E., \& Lattanzi, M. 1997, {\it ApJ}, {\bf 491}, 749

\bibitem{gro98}
Groom, D. E. 1998, {\it BAAS}, {\bf 30}, 1419

\bibitem{ham95}
Hamuy, M., Phillips, M. M., Maza, J., Suntzeff, N. B.,
   Schommer, R. A., \& Aviles, R. 1995, {\it AJ}, {\bf 109}, 1

\bibitem{ham96a}
Hamuy, M., Phillips, M. M., Maza, J., Suntzeff, N. B.,
   Schommer, R. A., \& Aviles, R. 1996a, {\it AJ}, {\bf 112}, 2391

\bibitem{ham96b}
Hamuy, M., Phillips, M. M., Maza, J., Suntzeff, N. B.,
   Schommer, R. A., \& Aviles, R. 1996b, {\it AJ}, {\bf 112}, 2398 

\bibitem{ham96c}
Hamuy, M., Phillips, M. M., Maza, J., Suntzeff, N. B.,
   Schommer, R. A., \& Aviles, R. 1996c, {\it AJ}, {\bf 112}, 2408 

\bibitem{ham00}
Hamuy, M., Trager, S. C., Pinto, P. A., Phillips, M. M.,
  Schommer, R. A., Ivanov, V., \& Suntzeff, N. B. 2000, {\it AJ}, 
  {\bf 120}, 1479

\bibitem{han00}
Hanany, S., et al. 2000, {\it ApJ}, {\bf 545}, L5

\bibitem{han98}
Hancock, S., Rocha, G., Lazenby, A. N., \& Guti\'{e}rrez, C. M. 1998, 
   {\it MNRAS}, {\bf 294}, L1

\bibitem{har00}
Hardin, D., et al. 2000, {\it A\&A}, {\bf 362}, 419

\bibitem{hat98}
Hatano, K., Branch, D., \& Deaton, J. 1998, {\it ApJ}, {\bf 502}, 177

\bibitem{hof98}
H\"{o}flich, P., Wheeler, J. C., \& Thielemann, F. K. 1998, {\it ApJ}, 
  {\bf 495}, 617

\bibitem{hol98}
Holz, D. E. 1998, {\it ApJ}, {\bf 506}, L1

\bibitem{holwal98}
Holz, D. E., \& Wald, R. 1998, {\it Phys. Rev. D}, {\bf 58}, 063501

\bibitem{hoy00}
Hoyle, F., Burbidge, G., \& Narlikar, J. V. 2000, {\it A Different
  Approach to Cosmology} (Cambridge: Cambridge Univ. Press)

\bibitem{iva00}
Ivanov, V. D., Hamuy, M., \& Pinto, P. A. 2000, {\it ApJ}, {\bf 542}, 
   588

\bibitem{jha02}
Jha, S. 2002, {\it Ph.D. thesis}, Harvard University

\bibitem{kai02}
Kaiser, N., et al. 2002, {\it BAAS}, {\bf 34}, 1304

\bibitem{kal03}
Kallosh, R., \& Linde, A. 2003, {\it J. Cosmology Astropart. Phys.}, 
   {\bf 2}, 2

\bibitem{kan98}
Kantowski, R. 1998, {\it ApJ}, {\bf 507}, 483

\bibitem{kan95}
Kantowski, R., Vaughan, T., \& Branch, D. 1995, {\it ApJ}, {\bf 447}, 35

\bibitem{kim96}
Kim, A., Goobar, A., \& Perlmutter, S. 1996, {\it PASP}, {\bf 108}, 190

\bibitem{kno03}
Knop, R., et al. 2003, {\it ApJ}, {\bf 598}, 102

\bibitem{kps}
Krisciunas, K., Phillips, M. M., \& Suntzeff, N. 2004, {\it ApJ}, 
   {\bf 602}, L81

\bibitem{kri01}
Krisciunas, K., et al. 2001, {\it AJ}, {\bf 122}, 1616

\bibitem{kri03}
Krisciunas, K., et al. 2003, {\it AJ}, {\bf 125}, 166

\bibitem{kri04}
Krisciunas, K., et al. 2004, {\it AJ}, {\bf 127}, 1664

\bibitem{lei01}
Leibundgut, B. 2001, {\it ARA\&A}, {\bf 39}, 67

\bibitem{lei93}
Leibundgut, B., et al. 1993, {\it AJ}, {\bf 105}, 301

\bibitem{lei96}
Leibundgut, B., et al. 1996, {\it ApJ}, {\bf 466}, L21

\bibitem{leo02a}
Leonard, D. C., et al. 2002a, {\it PASP}, {\bf 114}, 35 
   (erratum: {\bf 114}, 1291)

\bibitem{leo02b}
Leonard, D. C., et al. 2002b, {\it AJ}, {\bf 124}, 2490

\bibitem{wli03a}
Li, W., Filippenko, A. V., Chornock, R., \& Jha, S. 2003a,
  {\it ApJ}, {\bf 586}, L9

\bibitem{wli03c}
Li, W., Filippenko, A. V., Chornock, R., \& Jha, S. 2003c,
  {\it PASP}, {\bf 115}, 844

\bibitem{wli01}
Li, W., Filippenko, A. V., Treffers, R. R., Riess, A. G., Hu, J., 
   \& Qiu, Y. 2001b, {\it ApJ}, {\bf 546}, 734

\bibitem{wli00}
Li, W., et al. 2000, in {\it Cosmic Explosions}, ed. S. S.
  Holt \& W. W. Zhang (New York: AIP), 103

\bibitem{li01b}
Li, W., et al. 2001a, {\it PASP}, {\bf 113}, 1178

\bibitem{li03b}
Li, W., et al. 2003b, {\it PASP}, {\bf 115}, 453

\bibitem{lin98}
Lineweaver, C. H. 1998, {\it ApJ}, {\bf 505}, L69 

\bibitem{lin98}
Lineweaver, C. H., \& Barbosa, D. 1998, {\it ApJ}, {\bf 496}, 
   624

\bibitem{liv01}
Livio, M. 2001, in {\it Supernovae and Gamma-Ray Bursts: The Greatest
   Explosions since the Big Bang}, ed. K. Sahu, M. Livio, \& N.
   Panagia (Cambridge: Cambridge Univ. Press), 334

\bibitem{mad98}
Madau, P., Della Valle, M., \& Panagia, N. 1998, {\it MNRAS}, 
   {\bf 297}, L17

\bibitem{mao04}
Maoz, D., \& Gal-Yam, A. 2004, {\it MNRAS}, {\bf 347}, 951

\bibitem{mat01}
Matheson, T., Filippenko, A. V., Li, W., Leonard, D. C.,
  \& Shields, J. C. 2001, {\it AJ}, {\bf 121}, 1648

\bibitem{mat03}
Matheson, T., et al. 2003, {\it ApJ}, {\bf 599}, 394

\bibitem{matt01}
Matteucci, F., \& Recchi, S. 2001, {\it ApJ}, {\bf 558}, 351

\bibitem{mod01}
Modjaz, M., Li, W., Filippenko, A. V., King, J. Y.,
  Leonard, D. C., Matheson, T., Treffers, R. R., \& Riess, A. G. 2001,
   {\it PASP}, {\bf 113}, 308

\bibitem{nar97}
Narlikar, J. V., \& Arp, H. C. 1997, {\it ApJ}, {\bf 482}, L119

\bibitem{net02}
Netterfield, C. B., et al. 2002, {\it ApJ}, {\bf 571}, 604

\bibitem{nol04}
Nolta, M. R., et al. 2004, {\it ApJ}, {\bf 608}, 10

\bibitem{nom00}
Nomoto, K., Umeda, H., Hachisu, I., Kato, M., Kobayashi, C., \& 
   Tsujimoto, T. 2000, in {\it Type Ia Supernovae: Theory and Cosmology}, 
   ed. J. C. Niemeyer \& J. W. Truran (Cambridge: Cambridge Univ. 
   Press), 63

\bibitem{nor89}
Norgaard-Nielsen, H., et al. 1989, {\it Nature}, {\bf 339}, 523

\bibitem{nug00}
Nugent, P., 2001, in {\it Particle Physics and Cosmology:
   Second Tropical Workshop}, ed. J. F. Nieves (New York: AIP), 263

\bibitem{nug02}
Nugent, P., Kim, A., \& Perlmutter, S. 2002, {\it PASP}, {\bf 114}, 803

\bibitem{nug95}
Nugent, P., Phillips, M., Baron, E., Branch, D., \&
   Hauschildt, P. 1995, {\it ApJ}, {\bf 455}, L147

\bibitem{ost95}
Ostriker, J. P., \& Steinhardt, P. J. 1995, {\it Nature}, {\bf 377}, 600

\bibitem{osw96}
Oswalt, T. D., Smith, J. A., Wood, M. A., \& Hintzen, P. 1996,
   {\it Nature}, {\bf 382}, 692

\bibitem{pae02}
Paerels, F., Petric, A., Telis, G., \& Helfand, D. J. 2002, {\it BAAS}, 
   {\bf 34}, 1264

\bibitem{pai02}
Pain, R., et al. 1996, {\it ApJ}, {\bf 473}, 356

\bibitem{pai02a}
Pain, R., et al. 2002a, {\it ApJ}, {\bf 577}, 120

\bibitem{pai02b}
Pain, R., et al. 2002b, {\it BAAS}, {\bf 34}, 1169

\bibitem{par00}
Parodi, B. R., et al. 2000, {\it ApJ}, {\bf 540}, 634

\bibitem{pea01}
Peacock, J. A., et al. 2001, {\it Nature}, {\bf 410}, 169

\bibitem{per01}
Percival, W., et al. 2001, {\it MNRAS}, {\bf 327}, 1297

\bibitem{per95a}
Perlmutter, S., et al. 1995a, {\it ApJ}, {\bf 440}, L41

\bibitem{per95b}
Perlmutter, S., et al. 1995b, {\it IAUC 6270}

\bibitem{per97}
Perlmutter, S., et al. 1997, {\it ApJ}, {\bf 483}, 565

\bibitem{per98}
Perlmutter, S., et al. 1998, {\it Nature}, {\bf 391}, 51

\bibitem{per99}
Perlmutter, S., et al. 1999, {\it ApJ}, {\bf 517}, 565

\bibitem{phi93}
Phillips, M. M. 1993, {\it ApJ}, {\bf 413}, L105 

\bibitem{phi92}
Phillips, M. M., et al. 1992, {\it AJ}, {\bf 103}, 1632

\bibitem{phi99}
Phillips, M. M., et al. 1999, {\it AJ}, {\bf 118}, 1766

\bibitem{psk77}
Pskovskii, Yu. P. 1977, {\it Sov. Astron.}, {\bf 21}, 675

\bibitem{psk84}
Pskovskii, Yu. P. 1984, {\it Sov. Astron.}, {\bf 28}, 658

\bibitem{str03}
Reiss, D. 2000, {\it PhD thesis}, University of Washington

\bibitem{rie99b}
Riess, A. G., Filippenko, A. V., Li, W. D., \& Schmidt,
  B. P. 1999b, {\it AJ}, {\bf 118}, 2668

\bibitem{rie98a}
Riess, A. G., Nugent, P. E., Filippenko, A. V., Kirshner,
  R. P., \& Perlmutter, S. 1998a, {\it ApJ}, {\bf 504}, 935

\bibitem{rpk95}
Riess, A. G., Press, W. H., \& Kirshner, R. P. 1995, {\it ApJ}, 
  {\bf 438}, L17

\bibitem{rpk96a}
Riess, A. G., Press, W. H., \& Kirshner, R. P. 1996a, {\it ApJ}, 
   {\bf 473}, 88

\bibitem{rpk96b}
Riess, A. G., Press, W. H., \& Kirshner, R. P. 1996b, {\it ApJ}, 
   {\bf 473}, 588

\bibitem{rie97}
Riess, A. G., et al. 1997, {\it AJ}, {\bf 114}, 722

\bibitem{rie98b}
Riess, A. G., et al. 1998b, {\it AJ}, {\bf 116}, 1009

\bibitem{rie99a}
Riess, A. G., et al. 1999a, {\it AJ}, {\bf 117}, 707

\bibitem{rie99c}
Riess, A. G., et al. 1999c, {\it AJ}, {\bf 118}, 2675

\bibitem{rie00a}
Riess, A. G., et al. 2000, {\it ApJ}, {\bf 536}, 62

\bibitem{rie01}
Riess, A. G., et al. 2001, {\it ApJ}, {\bf 560}, 49

\bibitem{rie04a}
Riess, A. G., et al. 2004a, {\it ApJ}, {\bf 600}, L163

\bibitem{rie04b}
Riess, A. G., et al. 2004b, {\it ApJ}, {\bf 607}, 665

\bibitem{rui02}
Ruiz-Lapuente, P., et al. 1992, {\it ApJ}, {\bf 387}, L33

\bibitem{sah97}
Saha, A., et al. 1997, {\it ApJ}, {\bf 486}, 1 

\bibitem{sah01}
Saha, A., et al. 2001, {\it ApJ}, {\bf 562}, 314 

\bibitem{sal55}
Salpeter, E. E. 1955, {\it ApJ}, {\bf 121}, 161

\bibitem{san93}
Sandage, A., \& Tammann, G. A. 1993, {\it ApJ}, {\bf 415}, 1

\bibitem{san96}
Sandage, A., et al. 1996, {\it ApJ}, {\bf 460}, L15

\bibitem{sch98}
Schmidt, B. P., et al. 1998, {\it ApJ}, {\bf 507}, 46

\bibitem{scr04}
Scranton, R., et al. 2004, {\it Phys. Rev. Lett.}, submitted 
    (astro-ph/0307335)

\bibitem{smi02}
Smith, R. C., et al. 2002, {\it BAAS}, {\bf 34}, 1232

\bibitem{spe03}
Spergel, D. N., et al. 2003, {\it ApJS}, {\bf 148}, 175

\bibitem{str03}
Strolger, L. 2003, {\it PhD thesis}, University of Michigan

\bibitem{str04}
Strolger, L., et al. 2004, {\it ApJ}, {\bf 613}, 200

\bibitem{sul03}
Sullivan, M., et al. 2003, {\it MNRAS}, {\bf 340}, 1057

\bibitem{sun96}
Suntzeff, N. 1996, in {\it Supernovae and Supernova Remnants}, 
   ed. R. McCray \& Z. Wang (Cambridge: Cambridge Univ. Press), 41

\bibitem{sun96}
Suntzeff, N., et al. 1996, {\it IAUC 6490}

\bibitem{ton03}
Tonry, J. L., et al. 2003, {\it ApJ}, {\bf 594}, 1

\bibitem{tri97}
Tripp, R. 1997, {\it A\&A}, {\bf 325}, 871

\bibitem{tri98}
Tripp, R. 1998, {\it A\&A}, {\bf 331}, 815

\bibitem{tur96}
Turatto, M., et al. 1996, {\it MNRAS}, {\bf 283}, 1 

\bibitem{tur02}
Turner, M. S., \& Riess, A. G. 2002, {\it ApJ}, {\bf 569}, 18

\bibitem{tys01}
Tyson, J. A., \& Angel, R. 2001, in {\it The New Era of
   Wide Field Astronomy}, ed. R. Clowes, et al. 
   (San Francisco: ASP), 347

\bibitem{ume99}
Umeda, H., et al. 1999, {\it ApJ}, {\bf 522}, L43

\bibitem{van92}
van den Bergh, S., \& Pazder, J. 1992, {\it ApJ}, {\bf 390}, 34

\bibitem{vau95}
Vaughan, T. E., Branch, D., Miller, D. L., \& Perlmutter, S.
   1995, {\it ApJ}, {\bf 439}, 558

\bibitem{wam98}
Wambsganss, J., Cen, R., \& Ostriker, J. P. 1998, {\it ApJ}, {\bf 494}, 29

\bibitem{wan04}
Wang, Y., \& Tegmark, M. 2004, {\it Phys. Rev. Lett.}, {\bf 92}, 241302

\bibitem{wil03}
Williams, B., et al. 2003, {\it AJ}, {\bf 126}, 2608

\bibitem{wilson02}
Wilson, G., Cowie, L. L., Barger, A. J., \& Burke, D. J. 2002,
   {\it AJ}, {\bf 124}, 1258

\bibitem{wri01}
Wright, E. L. 2001, {\it BAAS}, {\bf 34}, 574

\bibitem{yun00}
Yungelson, L. R., \& Livio, M. 2000, {\it ApJ}, {\bf 528}, 108

\bibitem{zal97}
Zaldarriaga, M., Spergel, D. N., \& Seljak, U. 1997, {\it ApJ}, 
   {\bf 488}, 1

\end{chapthebibliography}
\end{document}